\documentclass[12pt]{article}
\usepackage{hyperref}
\usepackage{cite}
\usepackage{xcolor}
\usepackage{graphicx}
\usepackage{caption,subcaption}
\usepackage{amsmath}
\usepackage{amssymb}
\usepackage{xspace}
\usepackage{verbatim}
\usepackage{mathtools}
\usepackage{tikz-cd}
\usepackage{braket}
\usepackage{url}
\usepackage[normalem]{ulem}

\makeatletter
\@addtoreset{equation}{section}

\makeatletter
\renewcommand\section{\@startsection {section}{1}{\z@}%
                                   {-3.5ex \@plus -1ex \@minus -.2ex}
                                   {2.3ex \@plus.2ex}%
                                   {\normalfont\large\bfseries}}
\renewcommand\subsection{\@startsection{subsection}{2}{\z@}%
                                     {-3.25ex\@plus -1ex \@minus -.2ex}%
                                     {1.5ex \@plus .2ex}%
                                     {\normalfont\bfseries}}

\def\baselinestretch{1.2}
\newcommand{\be}{\begin{equation}}
\newcommand{\ee}{\end{equation}}
\newcommand{\beq}{\begin{eqnarray}}
\newcommand{\eeq}{\end{eqnarray}}

\newcommand{\gone}[1]{{}}

\definecolor{amber}{rgb}{1.0, 0.75, 0.0}

\begin{document}
\begin{titlepage}
\begin{flushright}
MAD-TH-18-09
\end{flushright}

\vfil

\begin{center}

{\bf \Large
 Topological Data Analysis for the String Landscape
}

\vfil

Alex Cole and Gary Shiu

\vfil

\texttt{acole4@wisc.edu, shiu@physics.wisc.edu}

\vfil

Department of Physics, University of Wisconsin, Madison, WI 53706, USA

\vfil
\end{center}

\begin{abstract}

\noindent Persistent homology computes the multiscale topology of a data set by using a sequence of discrete complexes. In this paper, we propose that persistent homology may be a useful tool for studying the structure of the landscape of string vacua. As a scaled-down version of the program, we use persistent homology to characterize distributions of Type IIB flux vacua on  moduli space for three examples: the rigid Calabi-Yau, a hypersurface in weighted projective space, and the symmetric six-torus $T^6=(T^2)^3$. These examples suggest that persistence pairing and multiparameter persistence contain useful information for characterization of the landscape  in addition to the usual information contained in standard persistent homology. We also study how restricting to special vacua with phenomenologically interesting low-energy properties affects the topology of a distribution.
\end{abstract}
\vspace{0.5in}

\end{titlepage}
\renewcommand{\baselinestretch}{1.05}  
\section{Introduction}
String theory appears to have an enormous number of vacua, making up what is called the \emph{string landscape}. The seminal work of \cite{Bousso:2000xa} pointed out that the presence of multiple fluxes leads to a discretuum of values for physical observables like the cosmological constant. A statistical approach to studying the landscape was advocated in \cite{Douglas:2003um}. It was further argued in 
\cite{Susskind:2003kw} that the existence and size of the landscape necessitated the use of anthropic arguments. Some subsequent work used statistics and explicit constructions to explore corners of the landscape and make arguments about stringy naturalness, especially with regard to the scale of supersymmetry breaking, the presence of various symmetries, and the cosmological constant \cite{Ashok:2003gk,Banks:2003es,Denef:2004ze,Douglas:2004zu,Denef:2004cf,Susskind:2004uv,Douglas:2004qg,Dine:2004is,Conlon:2004ds,Kallosh:2004yh,Marchesano:2004yn,Dine:2005yq,Acharya:2005ez,Dienes:2006ut,Gmeiner:2005vz,Douglas:2006xy}. Efforts were also made to propose alternatives to anthropics for vacuum selection in the landscape \cite{Firouzjahi:2004mx,Kobakhidze:2004gm,Davoudiasl:2006ax}. 
Despite the widespread belief in the landscape paradigm, skepticisms have also been 
raised \cite{Banks:2004xh,Banks:2012hx}. 

Now, with the advent of Big Data, we are perhaps well-positioned to ask previously out-of-reach questions about the landscape.\footnote{We should not confuse the Big in Big Data with the Big that characterizes the string theory landscape. Estimates suggest that the number of flux vacua for a typical geometry is around $10^{500}$ \cite{Ashok:2003gk,Denef:2004ze} and can be as large as $10^{272,000}$ \cite{Taylor:2015xtz}, with the number of geometries in a particular of F-theory ensemble bounded below by $\frac{4}{3}\times 2.96\times 10^{755}$ \cite{Halverson:2017ffz}. However, we might hope that studying larger subsets of the landscape than previously possible will provide hints toward a more complete picture.} Several recent papers have applied techniques from machine learning to the landscape and other data sets in string theory \cite{Abel:2014xta,He:2017aed,Krefl:2017yox,Ruehle:2017mzq,Carifio:2017bov,Wang:2018rkk,Bull:2018uow,Klaewer:2018sfl,Constantin:2018xkj,Mutter:2018sra,He:2018jtw}. One drawback of machine learning approaches, however, is their interpretability: while a neural network may be effective in classifying pictures of cats and dogs, its method for making this decision is generally opaque. If the goal is physical insight, a black box algorithm is not sufficient. For example, if one is classifying string theory vacua, one would like to understand {\it how} the classifier works.

In this paper, we propose studying distributions of string vacua using \emph{persistent homology} \cite{Edelsbrunner2002,zomorodian2005computing,carlsson2009topology,carlsson_2014} (see \cite{2018arXiv180903624P} for a recent historical review). Persistent homology is a technique from the field of Topological Data Analysis (TDA) that allows one to formalize the notion of the shape of a data set. Roughly, this is done by thickening each point in the data set and computing topological invariants at various stages of thickening. The \emph{persistence} of topologically nontrivial features like connected components, loops, voids, etc.\ throughout this thickening defines the multiscale topology of the data set. 

We argue that persistent homology is useful for characterizing distributions of string vacua, with the ultimate goal of understanding how structure in the distribution correlates with low-energy physics. One advantage that persistent homology has over machine learning techniques is its clear interpretation: we are simply computing topological invariants of simplicial complexes. Persistent homology has proven useful in diverse fields including sensor networks \cite{desilva2007}, image processing \cite{carlsson2008local}, bioinformatics \cite{Nicolau7265}, genomics \cite{chan2013topology}, protein structure \cite{Gameiro2015}, neuroscience \cite{2018arXiv180605167S}, cosmology \cite{2011MNRAS.414..350S,2011MNRAS.414..384S,Pranav:2016gwr,Cole:2017kve,Xu:2018xnz,Elbers:2018fus}, and many more. For example, persistent homology has been used to study the Cosmic Web of large-scale structure, composed of interlocking voids, filaments, and sheets at a variety of scales \cite{Pranav:2016gwr}. Void structure is also found in moduli space for the simplest string toy example, flux vacua on a rigid Calabi-Yau, studied in \cite{Denef:2004ze}. In the case of the Cosmic Web, it seems that at a scale of 300 Mpc, our galaxy resides within a void \cite{Keenan:2013mfa}. We might ask where our universe lives within the distribution of string theory vacua. As with the Cosmic Web, the answer to this question is necessarily multiscale. 

Persistent homology can be used to compare different distributions of string vacua (by summarizing what topological features are present and how they relate) and to understand where individual vacua reside within a distribution. We imagine that applying persistent homology to distributions of string vacua could be potentially useful for understanding vacuum selection (which becomes even more interesting when issues of computational complexity are considered \cite{Denef:2006ad,Denef:2017cxt,Bao:2017thx,Halverson:2018cio}) or tunneling in the landscape. Moreover, by choosing only special vacua with certain phenomenologically interesting properties and studying the restricted distributions with persistent homology, we may learn which low-energy properties of a string vacuum are simultaneously allowed, and where they live.

The structure of this paper is as follows: in Section \ref{sec:PH}, we outline how persistent homology describes the multiscale topology of a point cloud. In Section \ref{sec:flux}, we briefly review the construction of Type IIB flux vacua. We then apply persistent homology to distributions of Type IIB flux vacua on various backgrounds: the rigid Calabi-Yau, a hypersurface in weighted projective space, and the symmetric $T^6$. While these are simplified setups, they contain useful lessons for more general (and realistic) constructions. In Section \ref{sec:conc} we conclude.

It is worth noting that an early work \cite{Cirafici:2015pky} applied some of these techniques to string vacua, albeit in an incomplete fashion. In this paper we  
carried out a thorough study, with an eye towards 
a large-scale analysis of the landscape.
As we shall see, the new techniques and observables we developed have a wider applicability than the simple examples we studied.

\section{Persistent Homology}\label{sec:PH}
\begin{figure}
\centering
\includegraphics[width=0.5\textwidth]{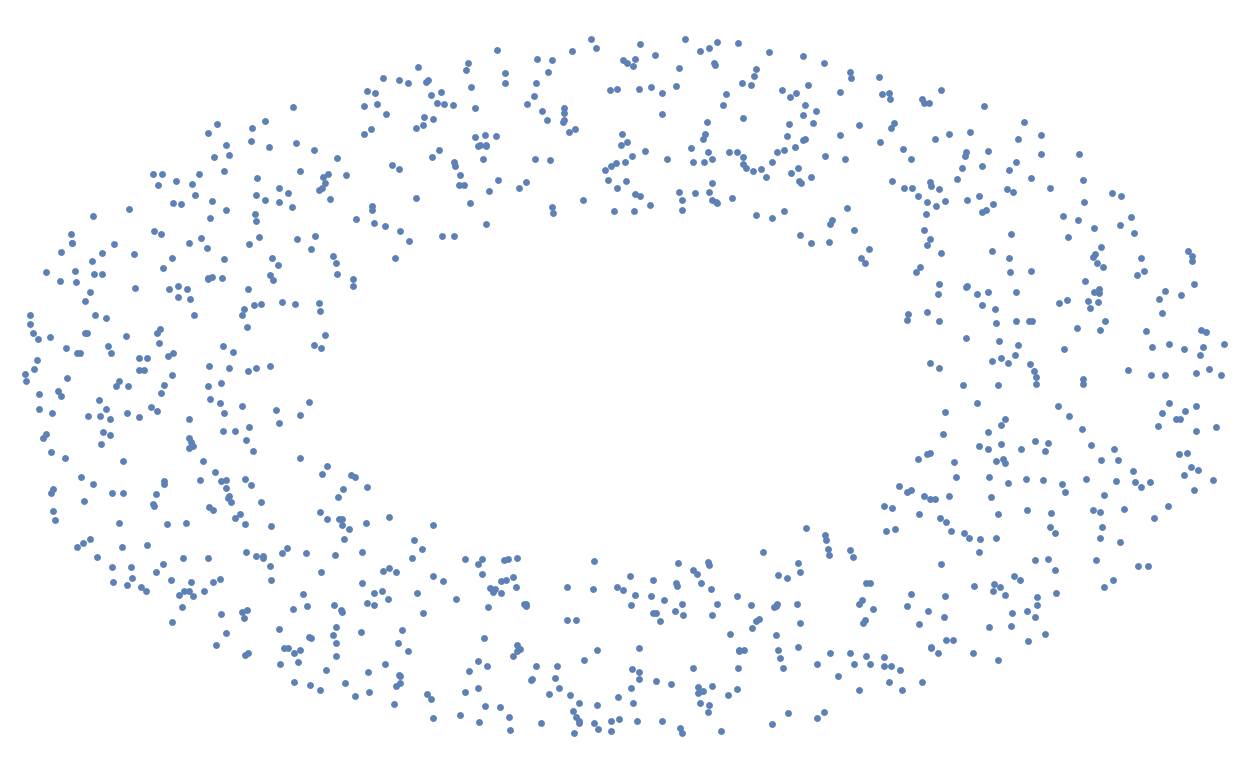}\caption{A point cloud generated by uniformly sampling points from an annulus. Persistent homology formalizes the presence of the hole in the middle.}\label{fig:ann}
\end{figure}

Persistent homology is a multiscale approach that can robustly characterize the shape of a data set. Though data sets are generally discrete, is it often the case that they contain patterns that are topological in nature. For example, if enough points are uniformly sampled from an annulus, the presence of a topologically nontrivial feature in the set of points is clearly visible (see Fig. \ref{fig:ann}). Persistent homology systematically describes the presence of such features by embedding the data set in a sequence of discrete complexes. Often each complex in the sequence can be thought of as a thickening of the points. In these cases, we can associate each complex in the sequence with a length scale given by the thickening. For each complex in the sequence, we are able to compute the number of topologically nontrivial features. Moreover, as we move through the sequence of complexes (to larger scales), we are able to track \emph{individual} topological features as they are created and destroyed. The \emph{persistence} of a topological feature in the sequence of complexes allows us to assign to it a notion of significance. The zeroth-order intuition is that long-lived features are real, robust aspects of the data, while short-lived features can be attributed to noise. (We will see in Sec. \ref{sec:rigid} that short-lived features can sometimes encode useful information about the data set's structure.)

In this section we briefly review simplicial complexes and persistent homology. We describe how witness complexes \cite{de2004topological} allow us to efficiently reconstruct the topology of a point cloud using a small sample of points. We then review persistence diagrams, which encode the output of a persistent homology computation. We emphasize that potentially useful information regarding persistence pairs is usually thrown out when making persistence diagrams. For a more in-depth discussion of persistent homology and computational topology in general, see \cite{zomorodian2005topology,edelsbrunner2010computational,carlsson_2014}.
\subsection{Simplicial complexes and persistent homology}
Consider a collection of points and pairwise distances between them. We call the collection of points a point cloud. We would like to formalize a notion of topology for the point cloud. To do this in a nontrivial way, we need to endow the point cloud with some extra structure. In this paper, we will embed the point cloud in a simplicial complex. More precisely, we will represent the point cloud with a sequence of simplicial complexes. We will then use persistent homology to track the creation and destruction of topological features as we move through the sequence of complexes. 

\begin{figure}
\centering
\includegraphics[width=0.5\textwidth]{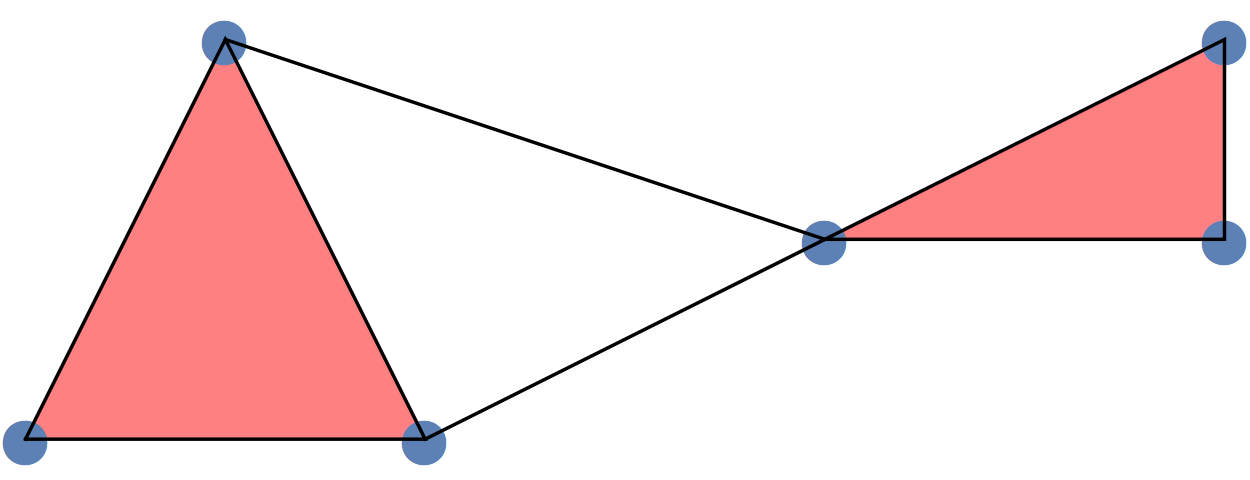}~~~~~~\includegraphics[width=0.4\textwidth]{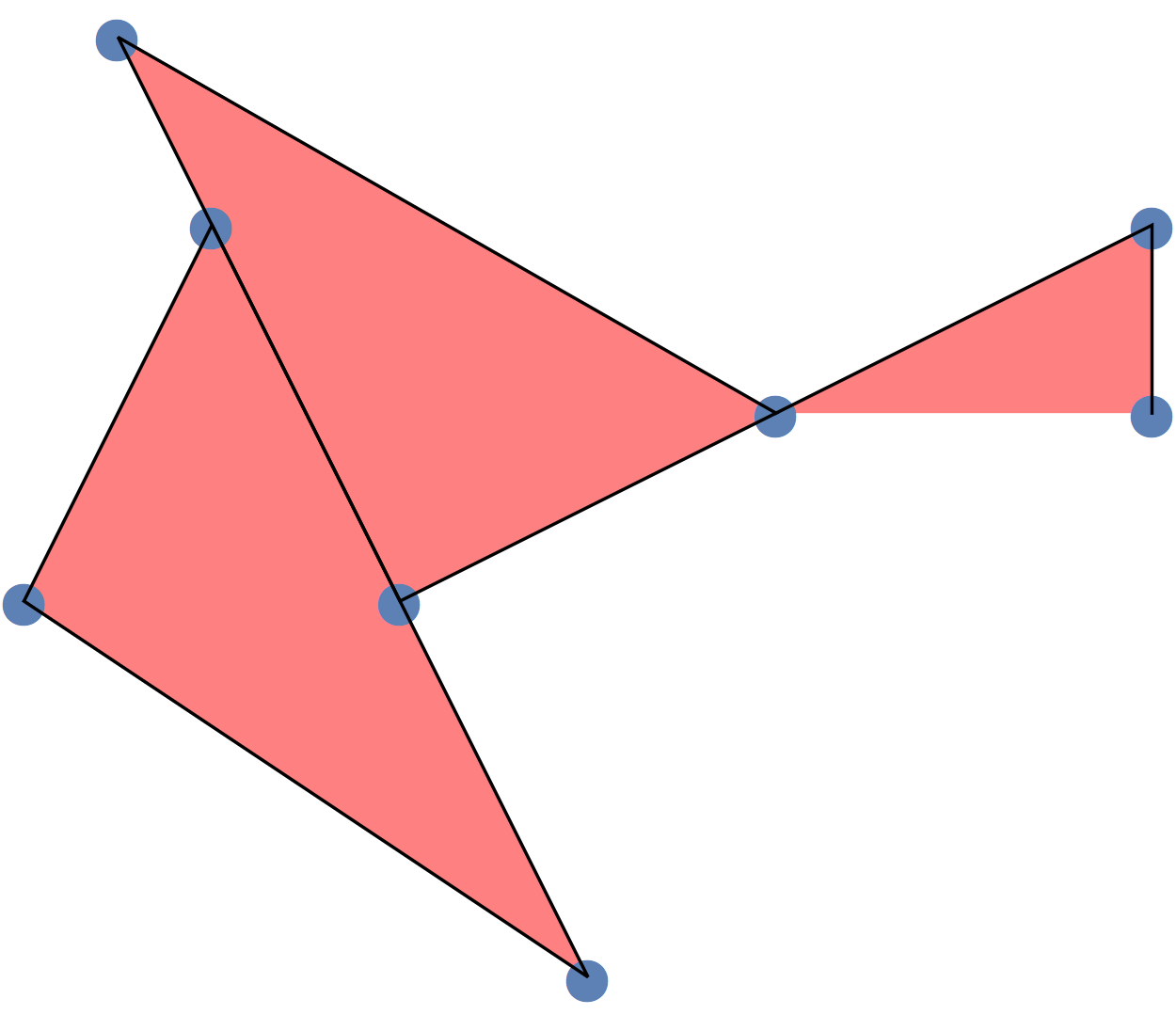}\caption{Left: a simplicial complex. There is a nontrivial 1-cycle given by the edges of the unshaded triangle. Right: the collection of simplices is not closed under taking faces or intersection of simplices.}\label{fig:comp}
\end{figure}
\subsubsection{Simplicial complexes and simplicial homology}
A simplicial complex is a collection of vertices (0-simplices), edges (1-simplices), triangles (2-simplices), etc.\ such that (i) all faces of a given simplex in the complex are contained in the complex (ii) the intersection of two simplices in the complex is contained in the complex. A simplicial complex and a collection of simplices that is not a simplicial complex are shown in Fig. \ref{fig:comp}. Given a simplicial complex, we can compute topological invariants. Consider some simplicial complex $S$. We define $k$-chains as collections of $k$-simplices in $S$. A $k$-chain may be formally represented as a sum
\begin{align}
	\sum_{i}a_i\sigma_i,\quad a_i\in\mathbb{Z}_2
\end{align}
where $i$ runs over $k$-simplices in $S$ and we use $\mathbb{Z}_2$ coefficients. The $k$-chains form a group under element-wise addition, which we denote $C_k$. There are two important subgroups of $C_k$. To define them, we first define the boundary operator $\partial_k:C_k\to C_{k-1}$. Writing a $k$-simplex in terms of its vertices as $[v_0,\dots,v_k]$, the action of the boundary operator is defined by 
\begin{align}
	\partial_k[v_0\dots v_k]=\sum_{j=0}^k [v_0,\dots,\hat{v}_j,\dots,v_k]
\end{align}
and linear extension. Here the hatted vertex is omitted. The boundary map $\partial_k$ is a homomorphism from $C_k$ to $C_{k-1}$. 

We call a $k$-chain $\sigma$ a $k$-cycle if its boundary vanishes, $\partial_k \sigma=0$. By linearity of the boundary map, the $k$-cycles form a subgroup of $C_k$, which we denote $Z_k$. In other words, $Z_k=\ker \partial_k$. Analogously, we define the group of $k$-boundaries, denoted $B_k$, as the image of the $(k+1)$ boundary map, $B_k=\textrm{im}~\partial_{k+1}$. In other words, $B_k$ is made up of $k$-simplices $\sigma$ such that, for some $(k+1)$-chain $\tau$, $\partial_{k+1}\tau=\sigma$.

Importantly, the boundary of a boundary is always empty, $\partial_{k}\partial_{k+1}=0$. This means that every $k$-boundary is also a $k$-cycle, so $B_k\subseteq Z_k$. However, the converse is not always true. For example, if a $k$-cycle wraps a $(k+1)$-dimensional hole, it cannot be written as the boundary of any $(k+1)$-chain in the complex. Thus we define the $k$-th homology group as $H_k(S,\mathbb{Z}_2)=Z_k/B_k$. In other words, elements of $H_k$ are $k$-cycles subject to the equivalence relation $\sigma\equiv \sigma+\tau$ for $\tau$ a $k$-boundary. The Betti numbers $b_k$ are the ranks of the homology groups. The 0-th Betti number $b_0$ counts the number of connected components and the higher Betti numbers $b_i$ count $(i+1)$-dimensional holes (by counting the cycles wrapping them). For example, the simplicial complex on the left side of Fig. \ref{fig:comp} has $b_0=1$ and $b_1=1$. For a general simplicial complex, the Betti numbers can be calculated by reducing the matrix representation of the boundary operator $\partial$ on the space of simplices in the complex \cite{edelsbrunner2010computational}.

So far we have only mentioned one simplicial complex. However, if we were to use just one simplicial complex to represent our data set, we would be asking for trouble. Given a point cloud, there are very many ways one might choose to form a simplicial complex. It is natural to associate each point with a vertex, but when it comes to connecting vertices with edges and other higher-dimensional simplices, we must make choices. With slightly different choices, one gets very different topological invariants. This is not a desirable outcome. To be confident about what we are learning from the data set, we need a more sophisticated approach.

\subsubsection{Persistent homology}

Persistent homology solves this problem of representational ambiguity by using the point cloud to construct a \emph{sequence} of simplicial complexes, called a filtration. Specifically, a filtration is a monotonic sequence of simplicial complexes $S_1\subset S_2\subset\dots\subset S_n$. As we move through the filtration, nontrivial cycles are created as simplices are added to the complex, and other cycles are made topologically trivial as they are filled in by simplices. Inclusion maps from $S_i$ to $S_{i+1}$ induce chain maps $C_k(S_{i})\to C_k(S_{i+1})$. Importantly, the corresponding homology maps $H_k(S_i)\to H_k(S_{i+1})$ are homomorphisms, since the inclusion maps commute with the boundary maps. This allows us to track \emph{individual} cycles in the homology as the complex grows. The output of a persistent homology computation is roughly a list of ordered pairs $(\nu_{\rm birth},\nu_{\rm death})$ detailing when in the filtration cycles of various dimensions are created and destroyed. (In fact, there is more information to harvest; see Sec. \ref{sec:pppd}.) This is stronger information than the Betti number curves $b_k(S_i)$, which merely count nontrivial cycles for each step in the filtration. Persistent homology resolves the problem of representational ambiguity, and has been proven stable (under a suitable metric) against perturbations to the point cloud \cite{cohen2005stability}. Moreover, moving through the filtration is often associated with some thickening scale, so persistent homology can rightfully be called a \emph{multiscale} technique.

\begin{figure}
\centering
\includegraphics[width=0.5\textwidth]{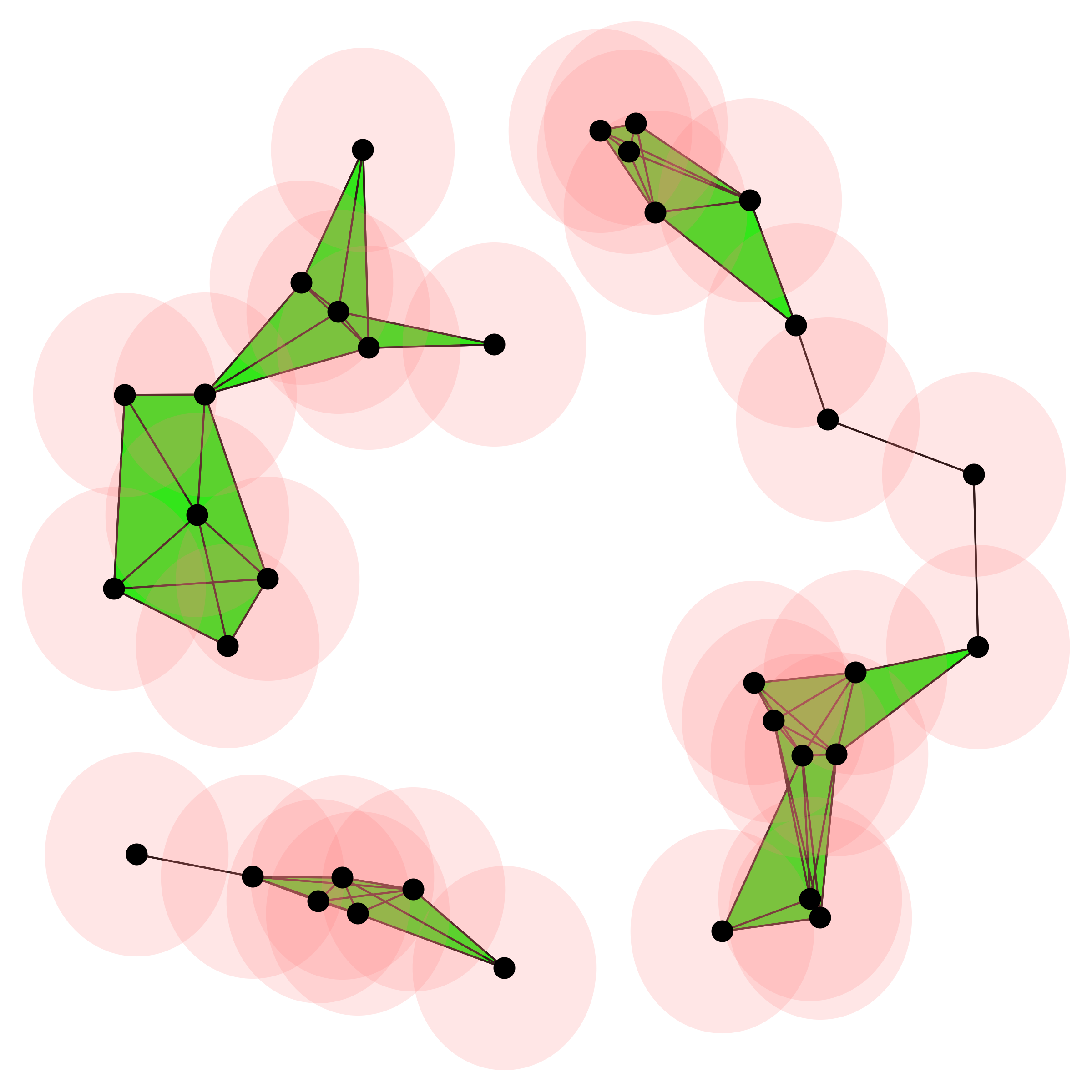}\caption{The Vietoris-Rips complex. Two vertices are connected by an edge if their disks (radius $r$) overlap. Higher-dimensional simplices like triangles are included if all of their faces are.}\label{fig:vr}
\end{figure}

As an example, consider the Vietoris-Rips complex. The vertex set of the Vietoris-Rips complex is the set of points in the cloud. Then, for $r>0$, we include an edge $[v_iv_j]$ if $d(v_i,v_j)<2r$. Higher-dimensional simplices are included if all of their faces are. An illustration is shown in Fig. \ref{fig:vr}. In this construction, the filtration parameter $r$ has a natural interpretation in terms of length. We are thickening each point to a ball of radius $r$ and connecting overlapping balls. As $r$ is increased, we consider the topology of the point cloud at larger and larger scales. At large enough $r$, the topology will become trivial, with $b_0=1$ and $b_i=0$ for $i>0$. This is true for any Vietoris-Rips complex.

\subsection{Witness complexes}
For our purposes, it will prove useful to use a more sophisticated construction than the Vietoris-Rips filtration. One issue with the Vietoris-Rips filtration is that it is very inefficient in terms of simplices. For example, a dense cluster of points leads to many edges and higher-dimensional simplices, but these are generally not arranged in a topologically interesting fashion. Computing and storing these simplices is wasteful.

One way to circumvent this problem is to use witness complexes\footnote{Another way to avoid the problem is to use $\alpha$-complexes \cite{Edelsbrunner:1993:UBD:160985.161139}, which intuitively treat the balls in a Vietoris-Rips type complex as bubbles that cannot overlap. However, this construction involves computing Delaunay triangulations, which is prohibitively costly in high-dimensional spaces. As we are eventually interested in string theory examples with many moduli, we use witness complexes.} \cite{de2004topological}. Witness complexes use a small subset of the point cloud as a landmark set, whose points form the vertex set of the complex. The presence of higher-dimensional simplices is determined by witness points, which can be any points in the data set. More precisely, let $L$ denote the set of landmark points, and $Z$ the full point cloud. Let $m_k(z)$ be the distance from $z\in Z$ to its $(k+1)$-th closest landmark point. Then for $k>0$ and vertices $l_i$, we include the $k$-simplex $[l_0l_1\dots l_k]$ in the witness complex $W(Z,L,r)$ if all of its faces are included and there exists a witness point $z\in Z$ such that
\begin{align}
	\max\{d(l_0,z),\dots,d(l_k,z)\}\leq r+m_k(z)
\end{align}
Here $r$ is the filtration parameter, analogous to the Vietoris-Rips filtration parameter. A simpler computation is the \emph{lazy} witness complex. For the lazy witness complex, one chooses $\nu\in \mathbb{N}$. If $\nu=0$, $m(z)$ is taken to be $0$. If $\nu>0$, $m(z)$ is the distance from $z$ to the $\nu$-th closest landmark point. The lazy witness complex is then constructed by taking the vertex set to be the landmark set and including an edge $[l_0l_1]$ if there is a witness $z\in Z$ such that
\begin{align}
	\max\{d(l_0,z),d(l_1,z)\}\leq r+m(z)
\end{align}
Then, as in the Vietoris-Rips construction, one includes a higher-dimensional simplex if all of its faces are included. This construction is called lazy because one only needs to compute the edges. For our purposes, we will generally use the lazy witness complex with $\nu=1$.

In general there are two ways to select a landmark set from a point cloud. One is to simply randomly choose points. The other is to use a maxmin algorithm, choosing the first point randomly and selecting subsequent points by maximizing the distance from the nearest landmark point. While maxmin gives more evenly spaced landmark points than a random selection, it tends to select outlier points; in point clouds with dense regions, one often obtains a more representative landmark set by using a random selector. We will generally use a maxmin selector. In Sec. \ref{sec:hyp} we will see that multiparameter persistence is also well suited for point clouds with dense regions.

\subsection{Persistence pairing and persistence diagrams}\label{sec:pppd}
Given a filtration, the persistent homology calculation is a matrix reduction of the boundary operator. Columns of the boundary operator represent individual simplices and are ordered by when a simplex is added to the filtration. See \cite{zomorodian2005computing,zomorodian2005topology,edelsbrunner2010computational} for details of the reduction algorithm. The reduced matrix encodes a \emph{persistence pairing} among simplices in the filtration. When a $k$-simplex is added to the filtration, it either creates a $k$-cycle or destroys a $(k-1)$-cycle. The persistence pairing links $k$-simplices and $(k+1)$-simplices (for all $k$ relevant to the complex) as persistence pairs, with the ($k+1$)-simplex destroying the cycle created by the $k$-simplex. (Some cycles may have infinite persistence, i.e.\ they are not destroyed by the end of the computed filtration, in which case there is no second simplex in the pair.) By looking at when the relevant simplices for a particular cycle were added to the filtration, one may assign to a cycle the filtration time of the cycle's birth as well as the filtration time of the cycle's death.

\begin{figure}
\centering
\includegraphics[width=0.5\textwidth]{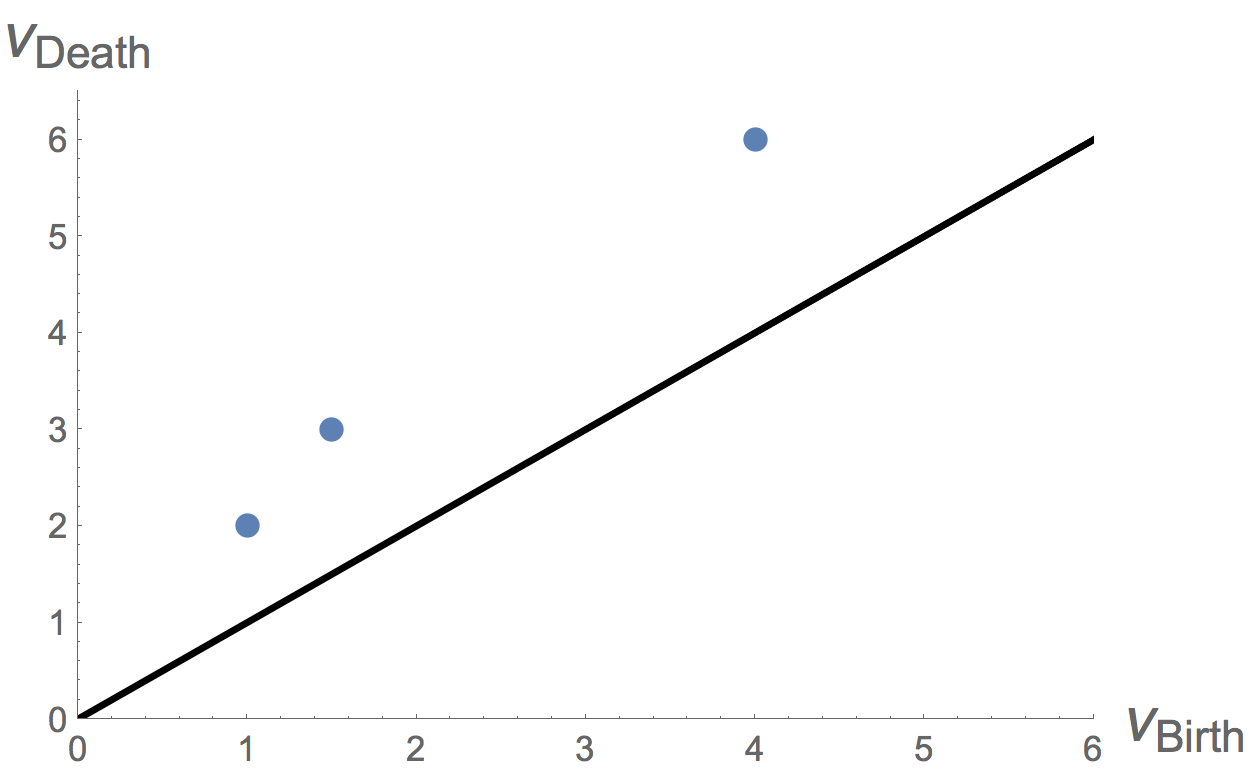}\includegraphics[width=0.5\textwidth]{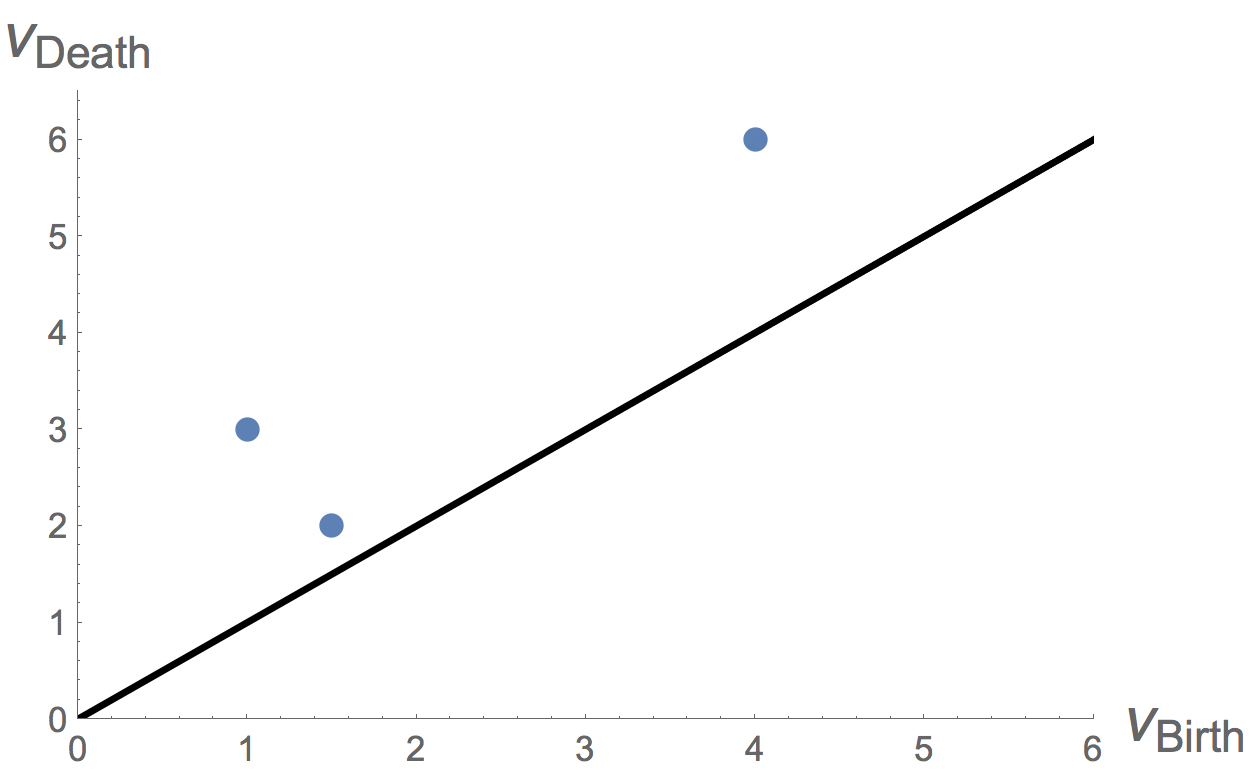}
\includegraphics[width=0.5\textwidth]{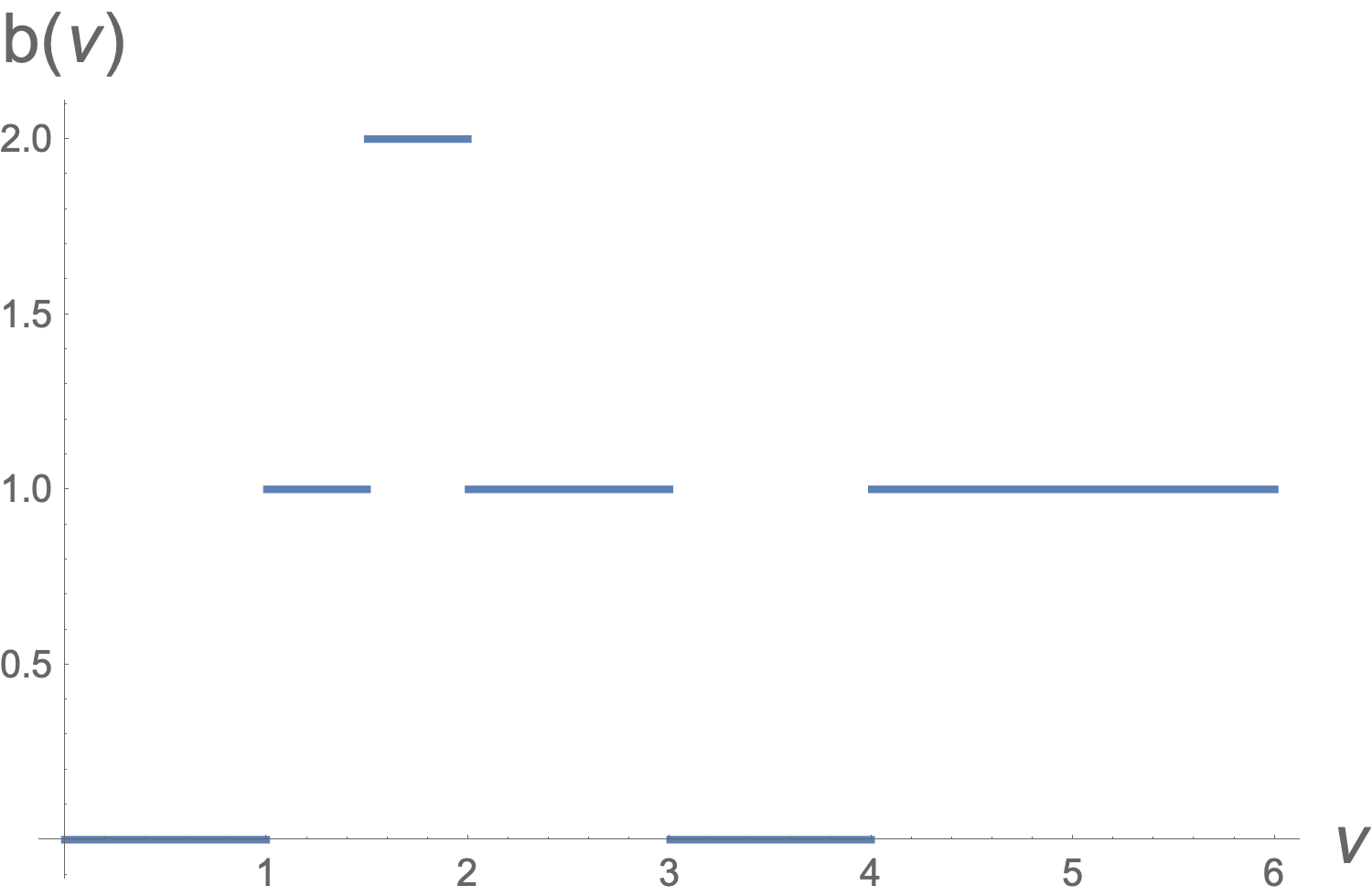}\caption{Two persistence diagrams. To calculate the Betti numbers at a given filtration time, one counts ``living'' cycles. Persistence diagrams contain more information that the Betti number curves. These two diagrams give rise to the same Betti number curve, shown below.}\label{fig:exPDs}
\end{figure}

Generally, there are two (equivalent) ways the output of a persistent homology calculation is represented. Barcodes are collections of horizontal lines, each starting at the birth time of a particular cycle and ending at that cycle's death time. One draws a barcode for each dimension of the homology. Persistence diagrams are scatter plots of the birth and death times $(\nu_{\rm birth},\nu_{\rm death})$ of individual cycles. Two examples of persistence diagrams (and their corresponding Betti number curves) are shown in Fig. \ref{fig:exPDs}.  Our experience is that relationships between cycles of different dimensions are easier to see using persistence diagrams. Since these relationships seem to matter for the distributions we are considering, we will use persistence diagrams. 

It should be noted that compressing persistent homology's output to barcodes and persistence diagrams actually erases some information about the structure of a data set. The persistence pairing computed by matrix reduction is used only to generate $(\nu_{\rm birth},\nu_{\rm death})$. However, the simplices appearing in the persistence pairing can encode more about the structure of a point cloud than just these two numbers. We will encounter a scenario in Sec. \ref{sec:rigid} where explicitly sorting through the persistence pairing and looking at overlapping ``destroying simplices'' allows us to recover more refined information about the point cloud than is represented in the persistence diagram.

\section{Flux Vacua}\label{sec:flux}
In this section, we consider type IIB string theory on Calabi-Yau orientifolds in the presence of background fluxes, as reviewed in \cite{Douglas:2006es,Grana:2005jc}.  In these setups, the axiodilaton and complex structure moduli are stabilized by 3-form fluxes threading the internal manifold. Flux quantization and tadpole cancellation give rise to a discretuum of vacua distributed over the moduli space \cite{Bousso:2000xa,Giddings:2001yu}. This discretuum generally features quite a bit of structure, well-suited for analysis via persistent homology. In particular, we will study the distribution of stabilized axiodilaton and complex structure moduli vevs using persistent homology. 

First we review the construction of flux vacua. We then study three examples using persistent homology: the rigid Calabi-Yau, a hypersurface in weighted projective space, and the symmetric $T^6$. Each example contains lessons that should be useful when applying persistent homology to more explicit and phenomenologically viable models. Specifically, we learn how persistence pairing, information that is often thrown out in a persistent homology calculation, can be useful. We also observe that a notion of \emph{multiparameter} perisistence should be used to characterize overdense regions in addition to the underdense regions one conventionally studies with persistent homology. Additionally, we study how different restrictions to special vacua (like those with enhanced symmetry or vanishing tree-level superpotential) can affect the topology of a distribution of string vacua.

As part of our simplified setup, we will study just the distribution of stabilized vevs for the axiodilaton and complex structure moduli, neglecting the distributions of K\"ahler moduli, open string moduli, and fluxes. For the K\"ahler moduli, we can appeal to a separation of scales, with the complex structure moduli and axiodilaton stabilized by fluxes and K\"ahler moduli stabilized by non-perturbative corrections to the no-scale models we consider. For further discussion of the motivation for neglecting the K\"ahler moduli, see \cite{Douglas:2003um,Ashok:2003gk,DeWolfe:2004ns}. We plan to eventually include these degrees of freedom, applying the lessons we learn from the following examples.
\subsection{Review}
We follow the conventions of \cite{DeWolfe:2004ns}. Consider a Calabi-Yau threefold $M$ with $h_{2,1}$ complex structure moduli. Take a symplectic basis $\{A^a,B_b\}$ for the $b_3=2h_{2,1}+2$ three-cycles, with $a,b=1,\dots,h_{2,1}+1$. We have dual cohomology elements $\alpha_a,\beta^b$ satisfying
\begin{align}
	\int_{A^a}\alpha_b=\delta^a_b,\quad \int_{B_b}\beta^a=-\delta^a_b,\quad\int_M \alpha_a\wedge \beta^b=\delta^b_a
\end{align}
From the unique holomorphic three-form $\Omega$, we have the periods $z^a\equiv \int_{A_a}\Omega,~\mathcal{G}_b\equiv\int_{B_b}\Omega$, which form the $b_3$-vector $\Pi(z)\equiv(\mathcal{G}_b,z^a)$. Additionally
\begin{align}
	\int_M\Omega\wedge\overline{\Omega}=\overline{z}^a\mathcal{G}_a-z^a\overline{\mathcal{G}}_a=-\Pi^\dag\cdot\Sigma\cdot\Pi
\end{align}
where we have introduced the symplectic matrix
\begin{align}
 	\Sigma=\left(\begin{matrix}0&1\\ -1&0\end{matrix}\right)
\end{align} 
whose entries are $(h_{2,1}+1)\times(h_{2,1}+1)$ matrices. The NSNS and RR 3-form fluxes are quantized and may be written in the $\alpha,\beta$ basis
\begin{align}
  F_3=-(2\pi)^2\alpha'(f_a\alpha_a+f_{a+h_{2,1}+1}\beta^a),\quad H_3=-(2\pi)^2\alpha'(h_a\alpha_a+h_{a+h_{2,1}+1}\beta^a)
\end{align}
where we have defined the integer-valued $b_3$-vectors $f$ and $h$. From now on we set $(2\pi)^2\alpha'=1$. The fluxes induce a superpotential for the complex structure moduli and axiodilaton $\phi\equiv C_0+ie^{-\varphi}$ \cite{Gukov:1999ya}
\begin{align}
  W=\int_M G_3\wedge \Omega(z)=(f-\phi h)\cdot \Pi(z)
\end{align}
where $G_3\equiv F_3-\phi H_3$. We are interested in vacua with vanishing F-terms
\begin{align}\label{eqn:fflat1}
  D_\phi W&=\frac{1}{\overline{\phi}-\phi}(f-\overline{\phi}h)\cdot \Pi(z)=0\\\label{eqn:fflat2}
  D_a W&= (f-\phi h)\cdot(\partial_a \Pi(z)+\Pi(z)\partial_a\mathcal{K})=0
\end{align}
where $D_aW\equiv\partial_a W+W\partial_a \mathcal{K}$, $a$ runs over complex structure moduli, and the K\"ahler potential truncated to the axiodilaton and complex structure moduli is
\begin{align}
  \mathcal{K}=-\log\left(i\int_M\Omega\wedge\overline{\Omega}\right)-\log\left(-i(\phi-\overline{\phi})\right)=-\log(-i\Pi^\dag\cdot\Sigma\cdot \Pi)-\log(-i(\phi-\overline{\phi}))
\end{align}
The F-flatness conditions (\ref{eqn:fflat1}) and (\ref{eqn:fflat2}) imply that the (3,0) and (1,2) parts of the fluxes vanish, so that $G_3$ is imaginary self-dual, $\star_6 G_3=iG_3$. 

The fluxes also contribute to the D3-brane charge
\begin{align}
  N_{\rm flux}=\int_M F_3\wedge H_3=f\cdot\Sigma\cdot h
\end{align}
For imaginary self-dual fluxes, we have that $N_{\rm flux}>0$. Therefore tadpole cancellation requires the presence of negative D3-brane charges. A fixed amount of negative charge is induced by orientifolding. If the orientifold can be viewed as arising from a fourfold compactification of F-theory \cite{Sen:1996vd}, the orientifold charge $L$ is proportional to the Euler character of the fourfold \cite{Sethi:1996es}. For cancellation, any difference $L-N_{\rm flux}$ can be made up by mobile D3-branes spanning the four-dimensional spacetime. For two of our three examples, we will not consider explicit orientifolds. Instead, we will take $L_{\rm max}$ as an adjustable parameter, so that 
\begin{align}
  0<N_{\rm flux}\leq L_{\rm max}
\end{align}
This is in line with the conventions of \cite{Ashok:2003gk,Denef:2004ze,DeWolfe:2004ns}. As we will see, $L_{\rm max}$ sets the scale at which we observe interesting structure in the moduli space distribution.

\subsection{Gauge-fixing}
We have gauge symmetries relating equivalent vacua that must be fixed. In general, our symmetry group is $\mathcal{G}=SL(2,\mathbb{Z})_\phi\times \Gamma$. Here $SL(2,\mathbb{Z})_\phi$ is the S-duality group from type IIB string theory and $\Gamma$ is the modular group of the complex structure moduli space.

Under $SL(2,\mathbb{Z})_\phi$, the axiodilaton and fluxes transform as
\begin{align}
  \phi\to\frac{a\phi+b}{c\phi+d},\quad \left(\begin{matrix}f\\h\end{matrix}\right)\to \left(\begin{matrix}a&b\\c&d\end{matrix}\right)\left(\begin{matrix}f\\h\end{matrix}\right),\quad a,b,c,d\in\mathbb{Z},\quad ad-bc=1
\end{align}
These transform solutions of (\ref{eqn:fflat1}), (\ref{eqn:fflat2}) to other solutions, and preserve $N_{\rm flux}$. They act as K\"ahler transformations on $W$ and $\mathcal{K}$
\begin{align}\label{eqn:kahler}
	W\to\Lambda W,\quad \mathcal{K}\to\mathcal{K}-\log\Lambda-\log\overline{\Lambda}
\end{align}
and are thus symmetries of $\mathcal{N}=1$ supergravity, with the scalar potential $V\equiv e^{\mathcal{K}}(|DW|^2-3|W|^2)$ manifestly invariant. 

We also have the complex structure modular group $\Gamma$. For our examples, under a transformation of the complex structure moduli $z^a\to z'^a$, the periods change as
\begin{align}
	\Pi(z^a)\to \Pi(z'^a)= \Lambda(z^a)M\cdot \Pi(z^a)
\end{align}
where $M$ is a symplectic matrix with integer entries that is independent of $z^a$. This transformation then induces a K\"ahler transformation (\ref{eqn:kahler}) and is thus a symmetry as long as the fluxes transform as
\begin{align}
	f\to f\cdot M^{-1},\quad h\to h\cdot M^{-1}
\end{align}
which also preserves $N_{\rm flux}$. Note that this transformation takes solutions to the F-flatness conditions to other solutions with moduli $\phi,z'^a$. Since we are ignoring the distribution of fluxes, the moral here is that we can apply modular transformations to the axiodilaton and complex structure moduli vevs independently and without keeping track of the fluxes.

\begin{figure}
\centering
\includegraphics[width=0.4\textwidth]{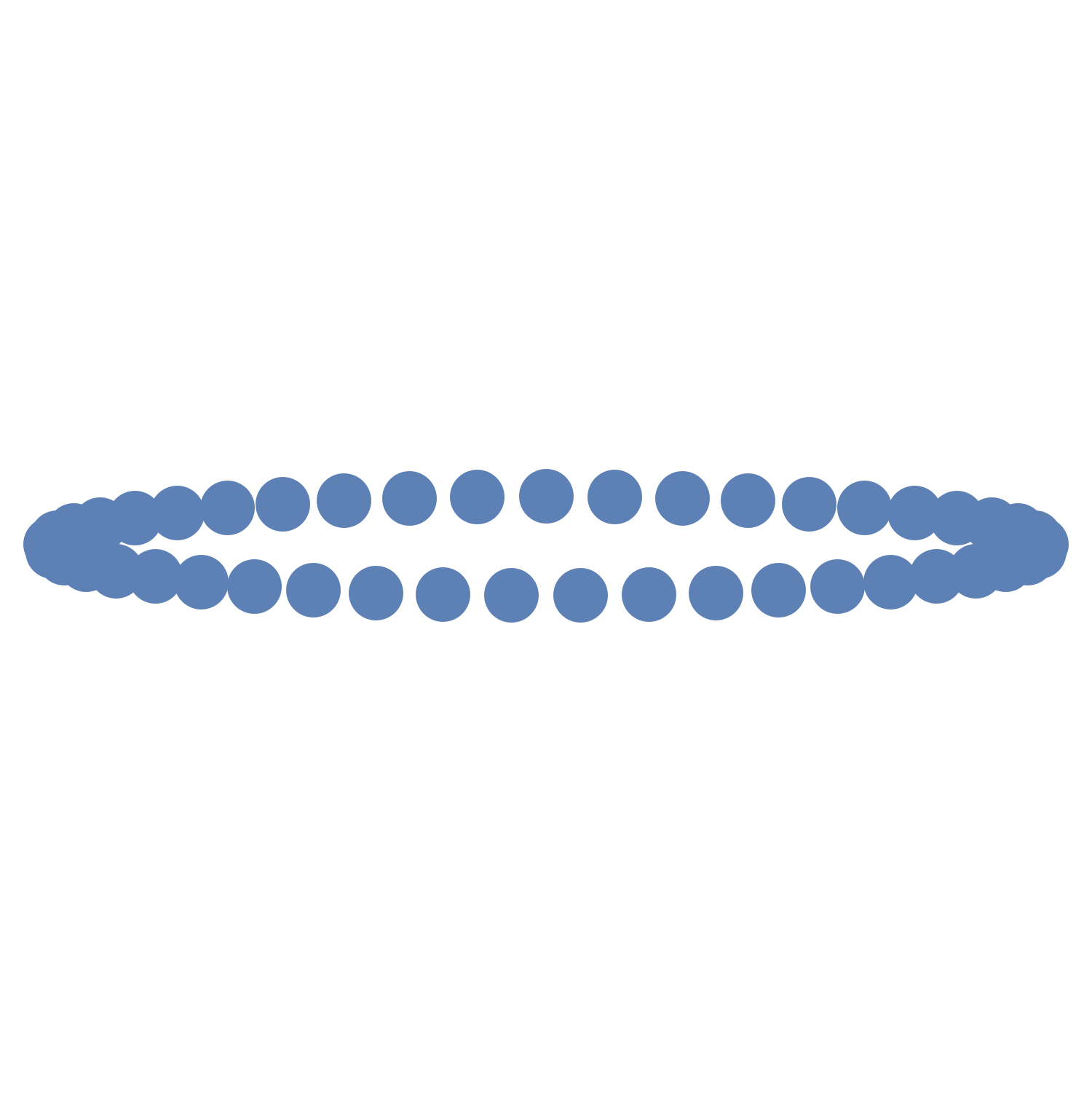}\quad\quad\quad\includegraphics[width=0.4\textwidth]{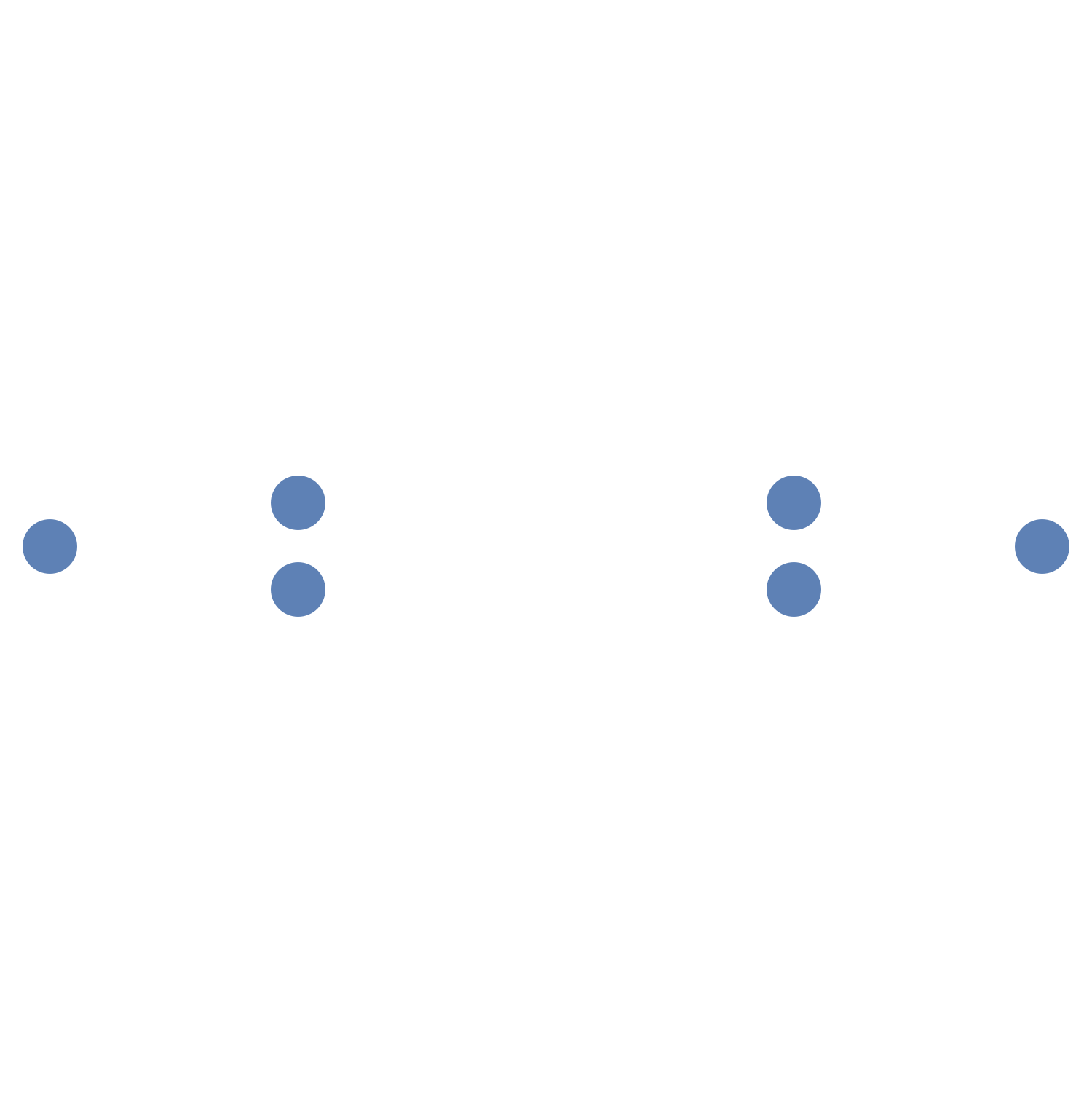}\caption{One effect of discreteness on topology. Left: a deformed well-sampled circle. Such features can accomodate more deformation than their poorly-sampled counterparts and still be recovered by persistent homology. Right: a deformed poorly-sampled circle. The deformation overtakes the characteristic distance between points, and the 1-cycle will be poorly recovered by persistent homology.}\label{fig:defsamp}
\end{figure}

Our gauge-fixing prescription will be to map each vacuum to a fundamental domain. One could also choose some gauge-image of the fundamental domain. In this case, continuity of the gauge map combined with the topological nature of our analysis would seem to suggest that the persistent homology would be stable under this operation (although the scales of certain features may change). There are interesting subtleties to this argument due to discreteness of the data set -- see Fig. \ref{fig:defsamp}. We have checked our results for stability under large gauge transformations.
\subsection{Enhanced symmetries and $W=0$ vacua}
Given the full set of flux vacua on some background with fixed $L_{\rm max}$, we are often interested in special vacua obeying certain phenomenologically desirable conditions. For example, in more involved setups including intersecting D-branes, one might enforce conditions such as appropriate ranks of gauge groups, number of generations of chiral matter, etc., along the lines of \cite{Gmeiner:2005vz,Douglas:2006xy}. One interesting question to ask is how the persistent homology of the distribution of vacua changes when one restricts to special vacua. Following \cite{DeWolfe:2004ns}, we will consider two types of special vacua: those with enhanced symmetries and those with vanishing tree-level superpotential.

Enhanced symmetries are low-energy symmetries (involving transformations of just the moduli) that descend from transformations of the moduli and the fluxes. The authors of \cite{DeWolfe:2004ns} consider enhanced symmetries that descend from the modular group as well as a complex conjugation transformation\footnote{In more complete setups, the complex conjugation transformation is related to spacetime $CP$. See \cite{DeWolfe:2004ns} and Sec. 16.5.1 of \cite{green1987superstring} for discussion of this point.}. One necessary condition for such a symmetry is that the moduli are invariant under the transformation \cite{DeWolfe:2004ns}. In other words, restricting onto vacua with enhanced symmetries involves restricting to fixed points in moduli space of the set of transformations. For $SL(2,\mathbb{Z})$, the fixed points are isolated in moduli space. For complex conjugation symmetries $\phi\to -\overline{\phi},z^a\to\pm\overline{z}^a$, the fixed points form a half-dimensional space, with each modulus restricted to its imaginary axis. Due to the dimensionalities of these subsets of moduli space, there is an upper bound on the dimensionalities of cycles in the restricted sets. For example, in a moduli space with $2n$ real dimensions, the region allowing vacua with a complex conjugation symmetry is $n$-dimensional. An $n$-dimensional region could in theory support an $n$-cycle, but the trivial topology of the $n$-dimensional plane means the highest cycle is an $(n-1)$-cycle. Restricting to these vacua erases cycles of dimension $n$ and higher that may be present in the full distribution of vacua.

Vacua with vanishing tree-level superpotential are interesting to study for a variety of reasons. By nonrenormalization theorems, one expects the condition $W=0$ is not corrected perturbatively. Assuming only nonperturbative (in $\alpha'$ or $g_s$) corrections, F-term SUSY breaking of these vacua leads to a plausibly naturally small cosmological constant \cite{DeWolfe:2004ns}. Moreover, it is known that there are deep connections between $W=0$ vacua and R-symmetries \cite{Nelson:1993nf}. In the $T^6$ example of Sec. \ref{sec:t6}, we will find that restricting to vacua with $W=0$ gives the distribution a richer topology, with more long-lived higher-dimensional features. This is possible in part because the condition $W=0$ itself does not select a subregion of the moduli space. In the continuous flux approximation and ignoring subtleties due to tadpole cancellation, there are $W=0$ vacua everywhere in the $T^6$ moduli space. The topology of the restricted set is then an effect of having quantized fluxes and imposing tadpole cancellation, which is also the reason for interesting structure in the full distribution of vacua. 

Understanding restrictions via their effects on the distribution's topology could be useful in a scaled-up problem where one wants to search for vacua that simultaneously satisfy several conditions, like certain number of generations, gauge groups, etc. In principle this could be done in a top-down fashion by analyzing large systems of equations, but in the limit of many conditions, a bottom-up tool like persistent homology seems potentially useful.

\subsection{Rigid Calabi-Yau}\label{sec:rigid}
Consider a rigid Calabi-Yau, with no complex structure moduli, studied in \cite{Ashok:2003gk,Denef:2004ze,DeWolfe:2004ns}. Since there are no complex structure moduli, $b_3=2$. We can write our symplectic basis for $H_3(M)$ as $\{A,B\}$. Take the periods of the holomorphic three-form $\Omega$ to be
\begin{align}
	\int_B\Omega=1,\quad \int_A\Omega=i
\end{align}
The superpotential is 
\begin{align}
	W=A\phi+B
\end{align}
where $A\equiv -h_1-ih_2,~B\equiv f_1+if_2$. The D3-brane charge induced by the fluxes is
\begin{align}
  N_{\rm flux}=f_1h_2-f_2h_1
\end{align}
and the vacuum equation is
\begin{align}
  D_\phi W=A\overline{\phi}+B=0
\end{align}
which is solved by
\begin{align}\label{eqn:rigid}
  \phi=-\overline{\left(\frac{B}{A}\right)}
\end{align}
We are interested in the distribution of flux vacua on the axiodilaton moduli space at fixed $L_{\rm max}$. To avoid repeat copies of individual vacua, we need to fix the gauge symmetry. For the rigid model, the entire modular group is $SL(2,\mathbb{Z})_\phi$. We can fix this gauge by either imposing conditions on the axiodilaton (i.e. restricting to a particular domain) or by imposing conditions on the fluxes. A natural condition is to restrict the axiodilaton to the $SL(2,\mathbb{Z})$ fundamental domain:
\begin{align}
  \phi\in\mathcal{F}_D=\left\{z\in \mathbb{C}_+:-\frac{1}{2}<\textrm{Re}(z)\leq \frac{1}{2},~|z|\geq 1\right\}
\end{align}
Alternatively, one can use $SL(2,\mathbb{Z})$ to impose conditions on the fluxes. For example, (as in \cite{Ashok:2003gk,DeWolfe:2004ns}), one may impose
\begin{align}\label{eqn:fluxcond}
  h_1=0,\quad 0\leq f_2<h_2
\end{align}
which entirely fixes $SL(2,\mathbb{Z})$. This strategy is particularly helpful for listing all vacua for a finite $L_{\rm max}$, or alternatively for counting vacua.

\begin{figure}\centering
\includegraphics[width=0.5\textwidth]{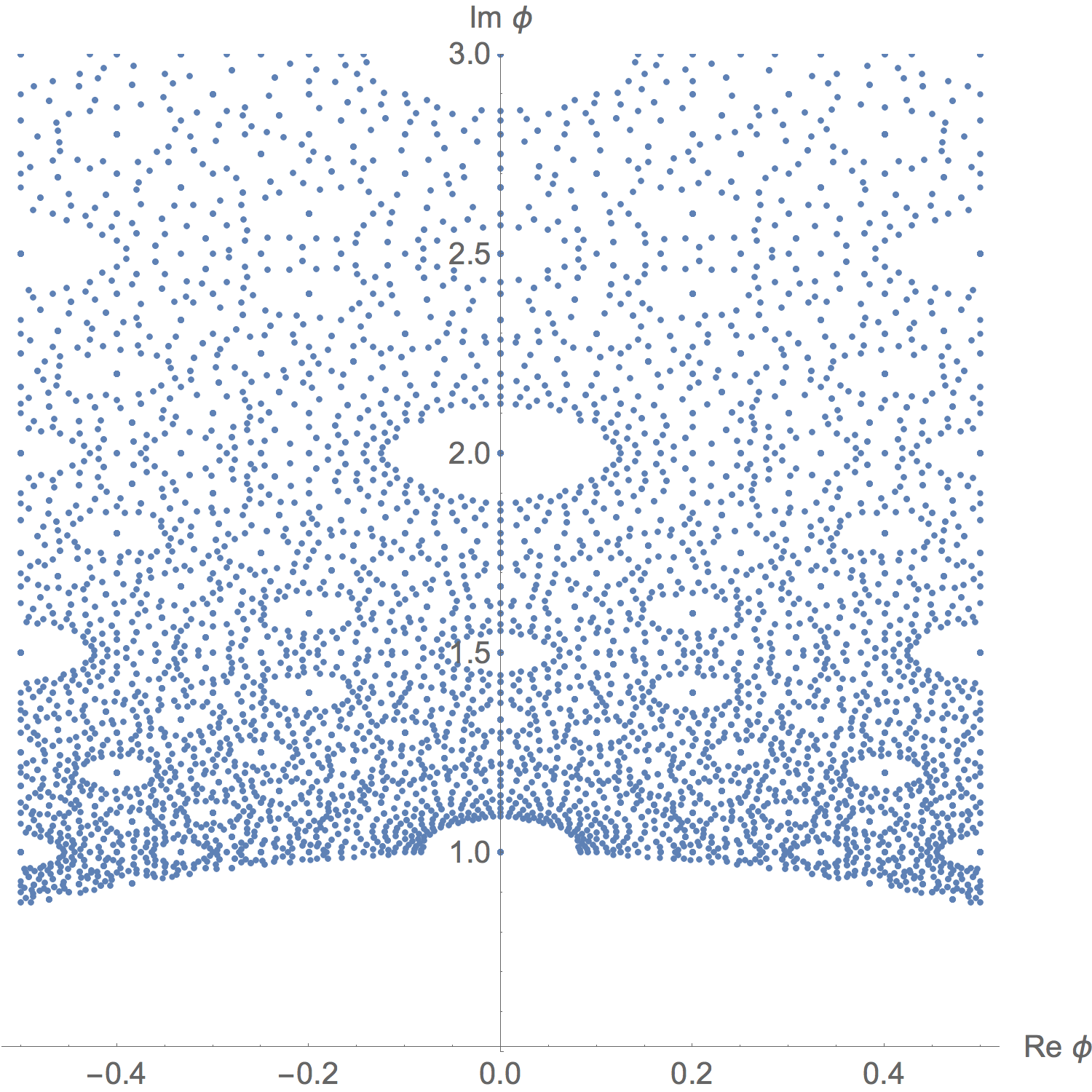}
\caption{Rigid Calabi-Yau vacua projected onto axiodilaton with $L_{\rm max}=150$. Each point in the diagram represents multiple inequivalent vacua. The relative sizes of the voids depend on number-theoretic aspects of the complex rational value of the axiodilaton at the center.}\label{fig:rigidAxio}
\end{figure}

To generate the distribution of flux vacua on the rigid Calabi-Yau, we use the flux conditions (\ref{eqn:fluxcond}) to list the vacua for finite $L_{\rm max}$. We then use $SL(2,\mathbb{Z})$ to map each vacuum to the axiodilaton fundamental domain. Despite the simplicity of the toy model, the distribution exhibits rich structure. In particular, projecting onto $\phi$, one observes voids with no vacua except for at accumulation points in their centers \cite{Denef:2004ze} (see Fig. \ref{fig:rigidAxio}). These voids can be understood as arising from the combination of flux quantization, tadpole cancellation, and the vacuum equation. The vacuum equation (\ref{eqn:rigid}) tells us that when we project onto $\phi$, we are introducing some degeneracy. There are multiple flux configurations that give the same stabilized value for $\phi$. These are inequivalent vacua, and represent different low energy theories. Specific values for $\phi$ correspond to 2-dimensional hyperplanes (intersecting with some gauge-fixing conditions) in the 4-dimensional flux space. Quantization of the fluxes forces these slopes to be rational. For example, $\phi=i\beta$ for rational $\beta$ corresponds to the hyperplane $f_1=\beta h_2,~f_2=-\beta h_1$. The hyperplanes intersect the origin in flux space, although no vacua are located there since the moduli are not stabilized for vanishing fluxes. Moving around in the axiodilaton moduli space corresponds to rotating these hyperplanes along two axes. However, tadpole cancellation with finite $L_{\rm max}$ means that most points in the axiodilaton moduli space are not represented in the distribution. Specifically, many hyperplanes fail to hit integer points in flux space before reaching the limits in flux space imposed by tadpole cancellation. These $\phi$ values are not present in the resulting distribution. 

Given a $\phi$ that is present in the distribution, its nearest neighbors are found by rotating the corresponding hyperplane in flux space until it hits a point with integer fluxes within the bounds imposed by tadpole cancellation. For concreteness, take $\phi=i\beta$. If such a $\phi$ is present one of the corresponding vacua takes the form $\frac{if_1}{h_2}$. Without loss of generality assume that $f_1$ and $h_2$ have no common factors and that both are positive. The nearest vacuum on the imaginary axis takes the form $i\left(\frac{f_1}{h_2}-\frac{a}{b}\right)$ with $a,b>0$. (One should use a plus sign for $f_1=h_2$ to stay in the fundamental domain.) To find the nearest neighbor, one minimizes $\frac{a}{b}$ over the naturals subject to the constraint $N_{\rm flux}=b^2f_1h_2-abh_2^2\leq L_{\rm max}$.

As an example, for $L_{\rm max}=150$, the nearest neighbors of $2i$ on the imaginary axis are $\frac{16\pm1}{8}i$ , while the nearest neighbor of $\frac{3}{2}i$ is $\frac{21-1}{14}i$. (Sometimes the plus sign takes one outside the tadpole bound, so that the nearest neighbor above is farther away than the nearest neighbor below.) Similar arguments to the above apply for nearest neighbors in other directions. Since $b$ scales as $\sqrt{L_{\rm max}}$ in this discussion, we also come to understand the previously known fact that the voids shrink as $\sqrt{1/L_{\rm max}}$ as $L_{\rm max}$ is increased \cite{Denef:2004ze,DeWolfe:2004ns}.

\begin{figure}
\centering
\includegraphics[width=0.3\textwidth]{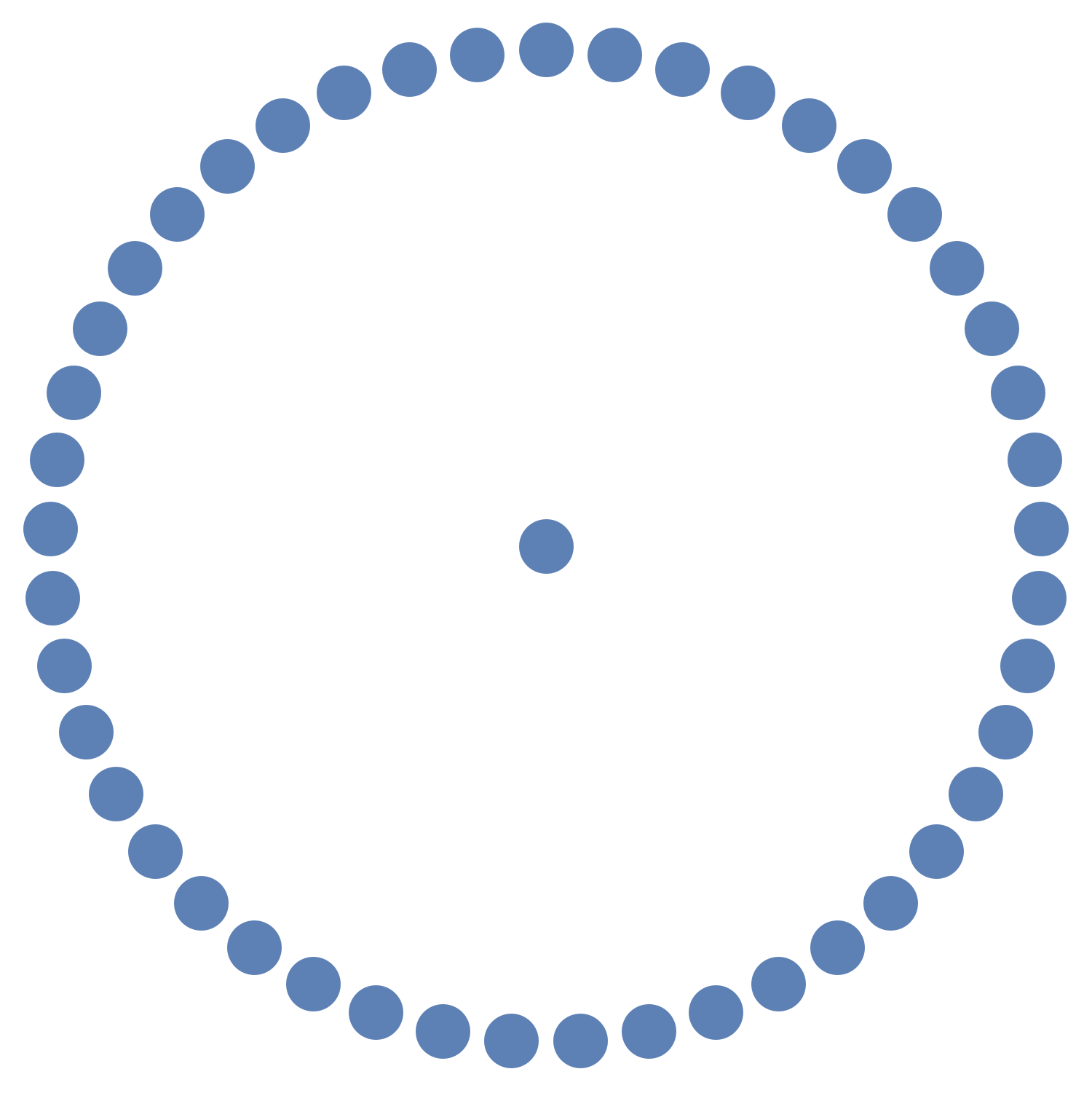}\quad\quad\quad\includegraphics[width=0.3\textwidth]{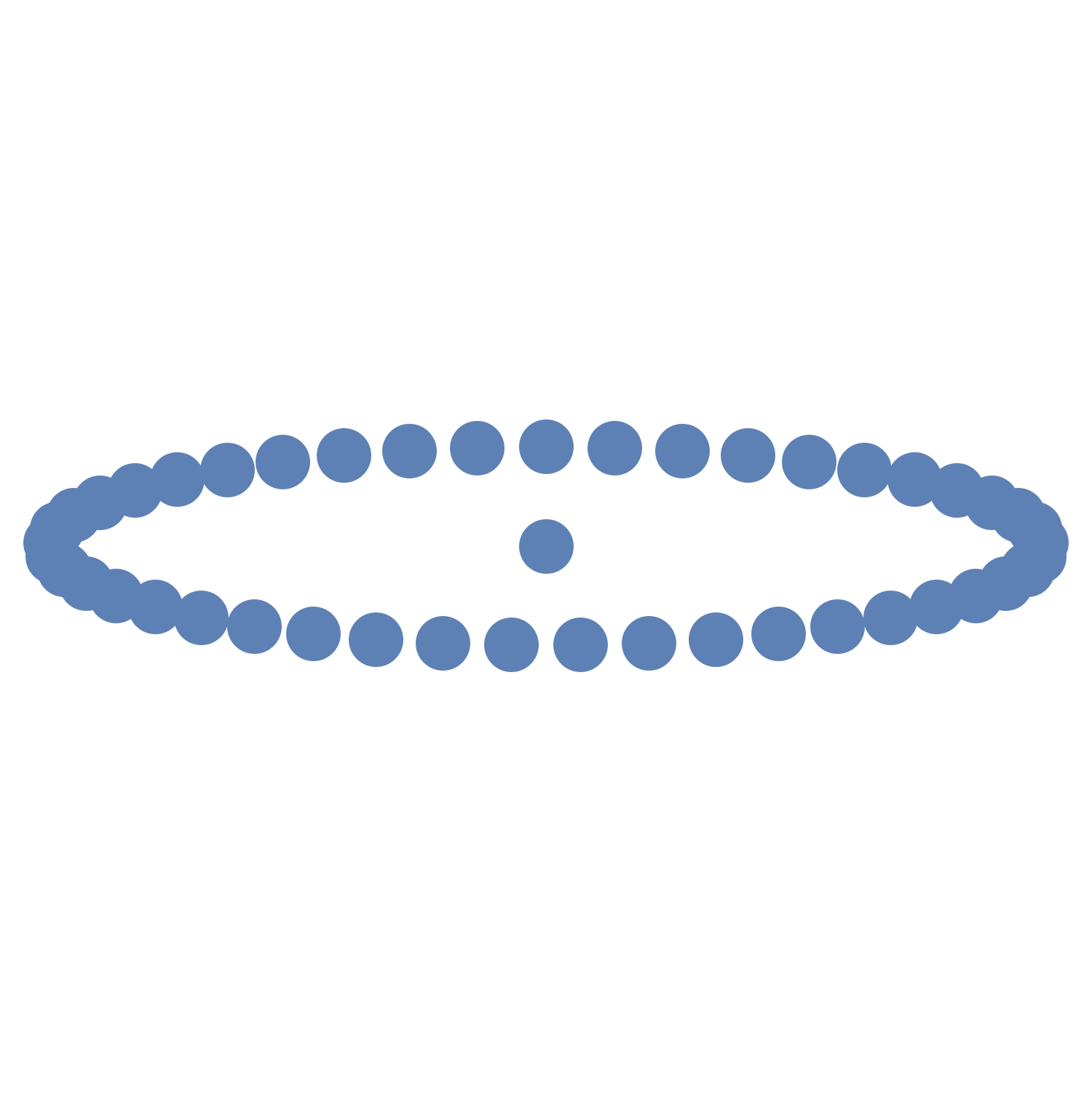}\caption{Left: a circular void with a vacuum in the middle. The long-lived 0-cycle will die at the same time as the long-lived 1-cycle. Right: an oblique void with a vacuum in the middle. The 0-cycle will die before the 1-cycle dies, and an extra short-lived 1-cycle will form.}\label{fig:defVoids}
\end{figure}

In terms of persistent homology, large voids should correspond to long-lived 1-cycles. Moreover, we expect the presence of vacua at the center of a void to have a specific signature in the persistence diagram. That is, the death of the 1-cycle corresponding to the void will be \emph{correlated} with the death of a 0-cycle. In the case of a perfectly symmetric void, the 1-cycle and 0-cycle will die at exactly the same time. For a more oblique void, the 0-cycle will die first, possibly with the formation of a short-lived 1-cycle (see Fig. \ref{fig:defVoids}). 
\begin{figure}\centering
  \includegraphics[width=0.5\textwidth]{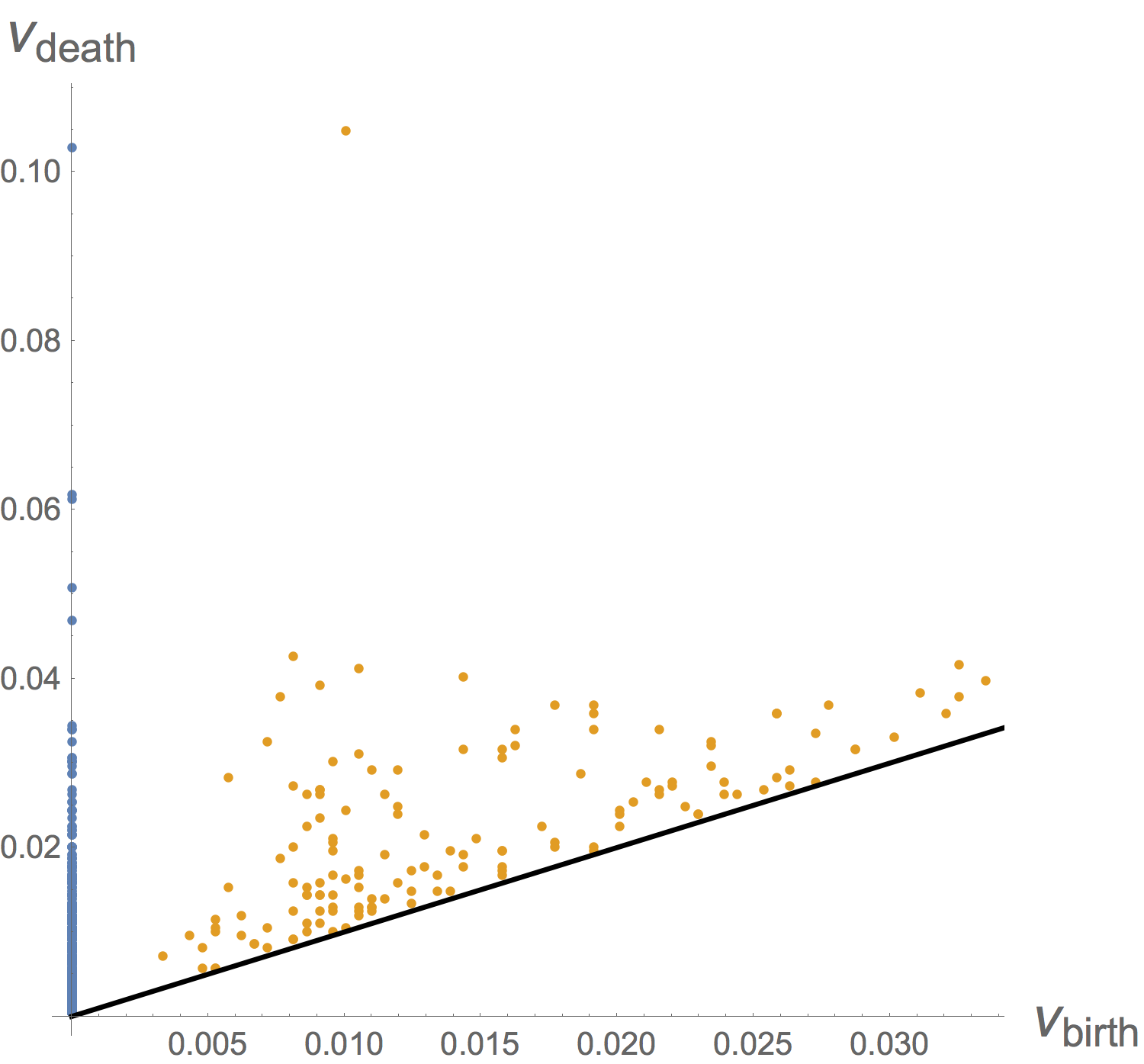}\includegraphics[width=0.5\textwidth]{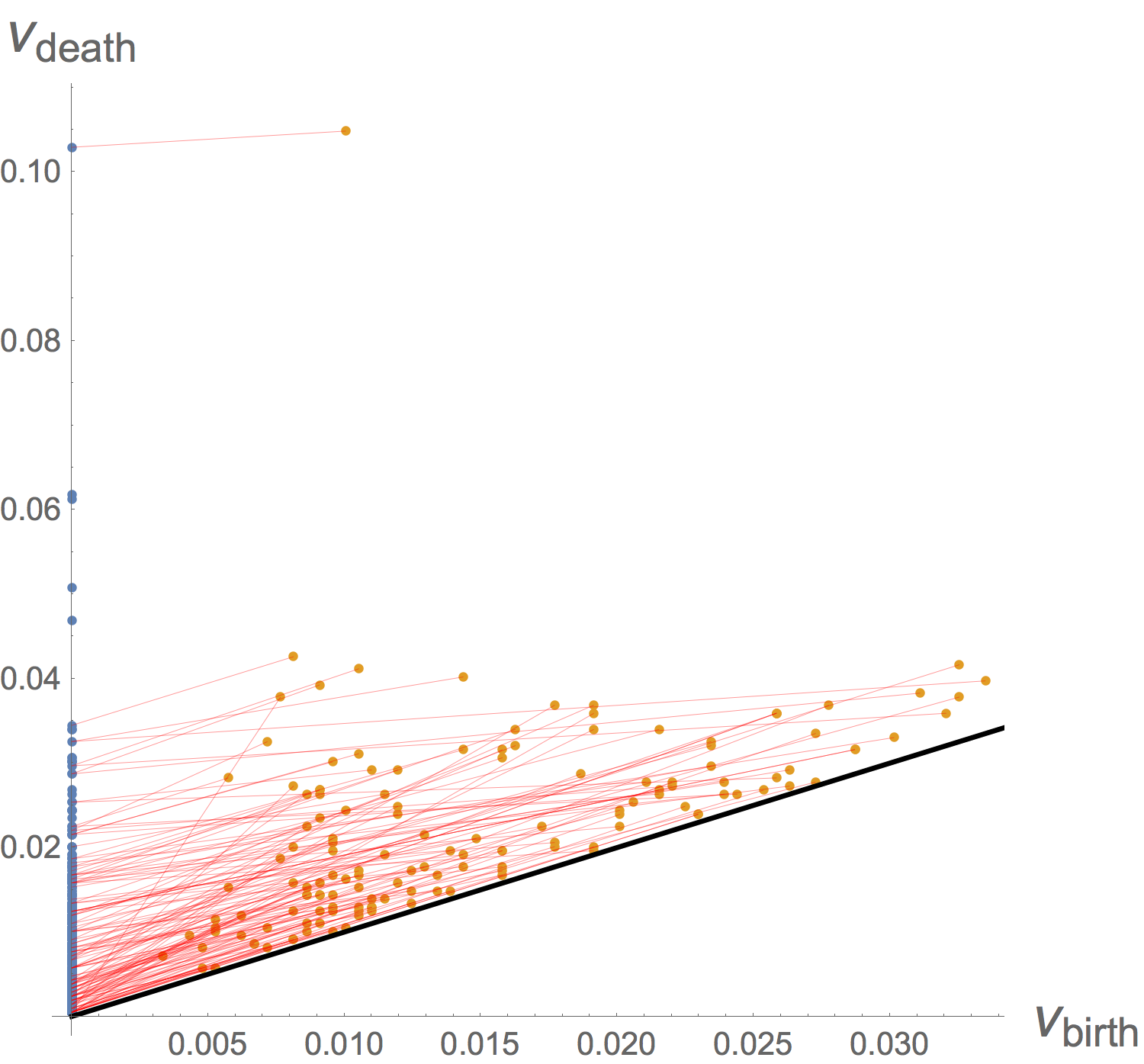}
  \caption{Left: persistence diagram for rigid Calabi-Yau flux vacua projected onto axiodilaton with $L_{\rm max}=150$, using a lazy witness complex with 700 landmark points. The filtration has 1,140,182 simplices. The orange points represent 1-cycles and the blue points represent 0-cycles. We observe many long-lived 1-cycles, corresponding to voids in the distribution. There are also correlations in the deaths of long-lived 1-cycles and 0-cycles, corresponding to the vacua in the centers of the voids. Right: making use of the information contained in persistence pairing, we can further investigate the correlated deaths. 0-cycles are destroyed by the addition of 1-simplices, and 1-cycles are destroyed by the addition of 2-simplices. We link with a red line 0-cycles and 1-cycles whose destroying simplices overlap.}\label{fig:rigidPDs}
\end{figure}

Persistence diagrams for a subregion of the rigid Calabi-Yau axiodilaton distribution are shown in Fig. \ref{fig:rigidPDs}. We observe the presence of many 1-cycles, some of which are relatively long-lived. The longer-lived 1-cycles correspond to the larger voids in Fig. \ref{fig:rigidAxio}. We also observe the expected correlated deaths of 1-cycles and 0-cycles. While many correlated deaths occur at the exact same filtration time, some of the voids slightly outlive their isolated center vacua, including the longest-lived void. We also note several 0-cycles with $\nu_{\rm death}\sim 0.06$ that do not seem to correlate with any 1-cycles. In fact, these 0-cycles correspond to vacua that \emph{would} be at the centers of voids, but whose voids are cut off by the boundary of $\mathcal{F}_D$. This boundary prevents the voids from being recognized topologically, other than the late deaths of the 0-cycles corresponding to isolated interior vacua.
\subsubsection{Persistence pairing}
We should note that while the correlated deaths are suggestive, they are not sufficient to recover the isolated vacua in the centers of the voids. Instead, one must turn to the persistence pairing output by the persistent homology algorithm. This pairing tells us which simplex causes the death of a particular $p$-cycle. For example, the death of the 0-cycle corresponding to an isolated vacuum in the center of a void is caused by the addition of an edge connecting that vacuum to a point on the edge of the void. This same edge is contained in the triangles that fill in the void, causing the 1-cycle's death. Persistence pairing is not shown in persistence diagrams\footnote{In part, this is because we plot only cycles in persistence diagrams, while ``destroying simplices'' are, at the filtration time they are added to the complex, merely simplices.} and represents finer-grained information. However, noting the correlated deaths from the previous section, we can use persistence pairing to connect 1-cycles and 0-cycles whose ``destroying simplices'' overlap. These connections are also shown in Fig. \ref{fig:rigidPDs}. They confirm the isolated vaccum-void structure, which we might have only suspected from the persistence diagram.
\subsubsection{Special vacua}
\begin{figure}
\centering
\includegraphics[width=0.5\textwidth]{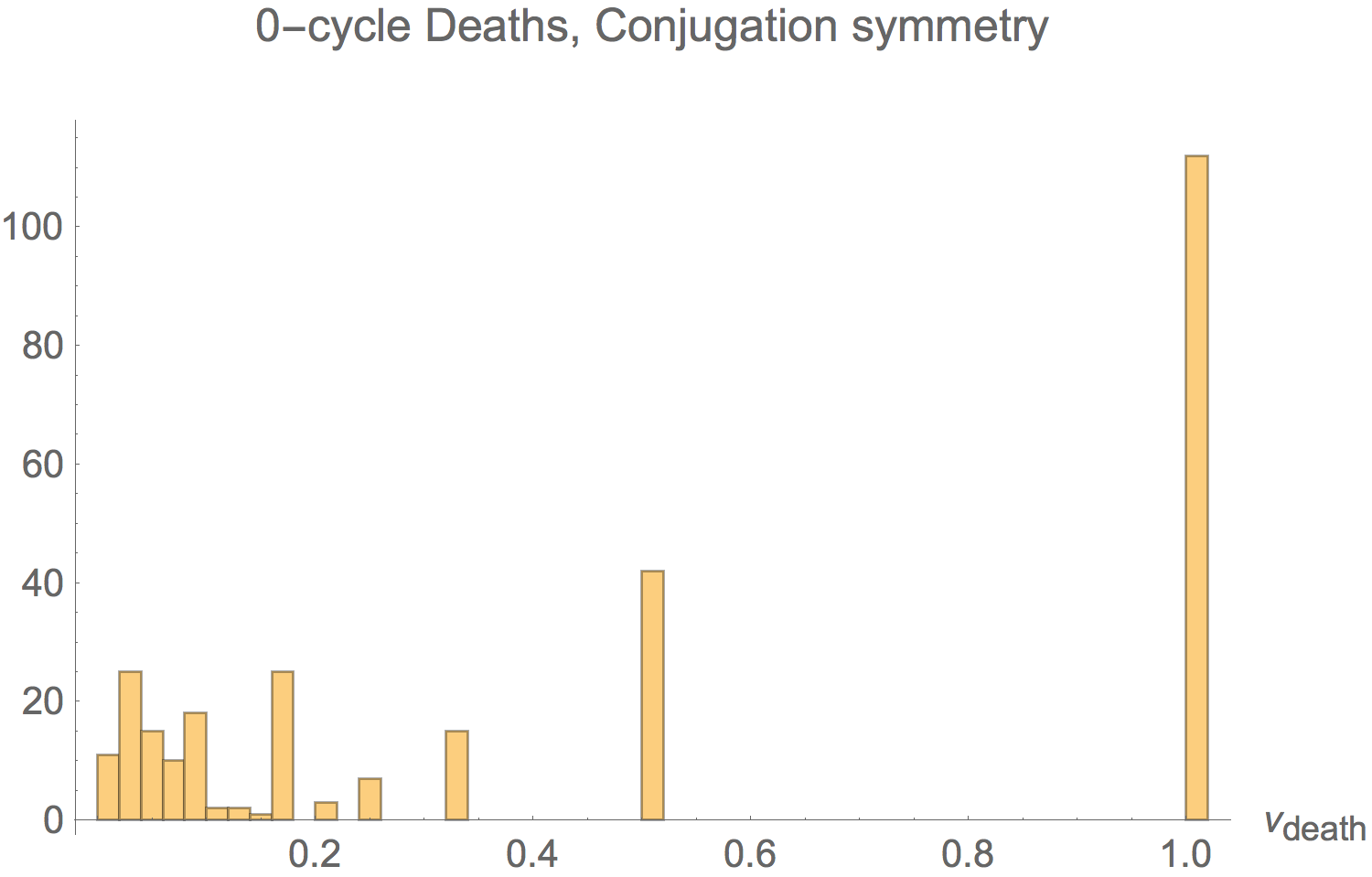}\includegraphics[width=0.5\textwidth]{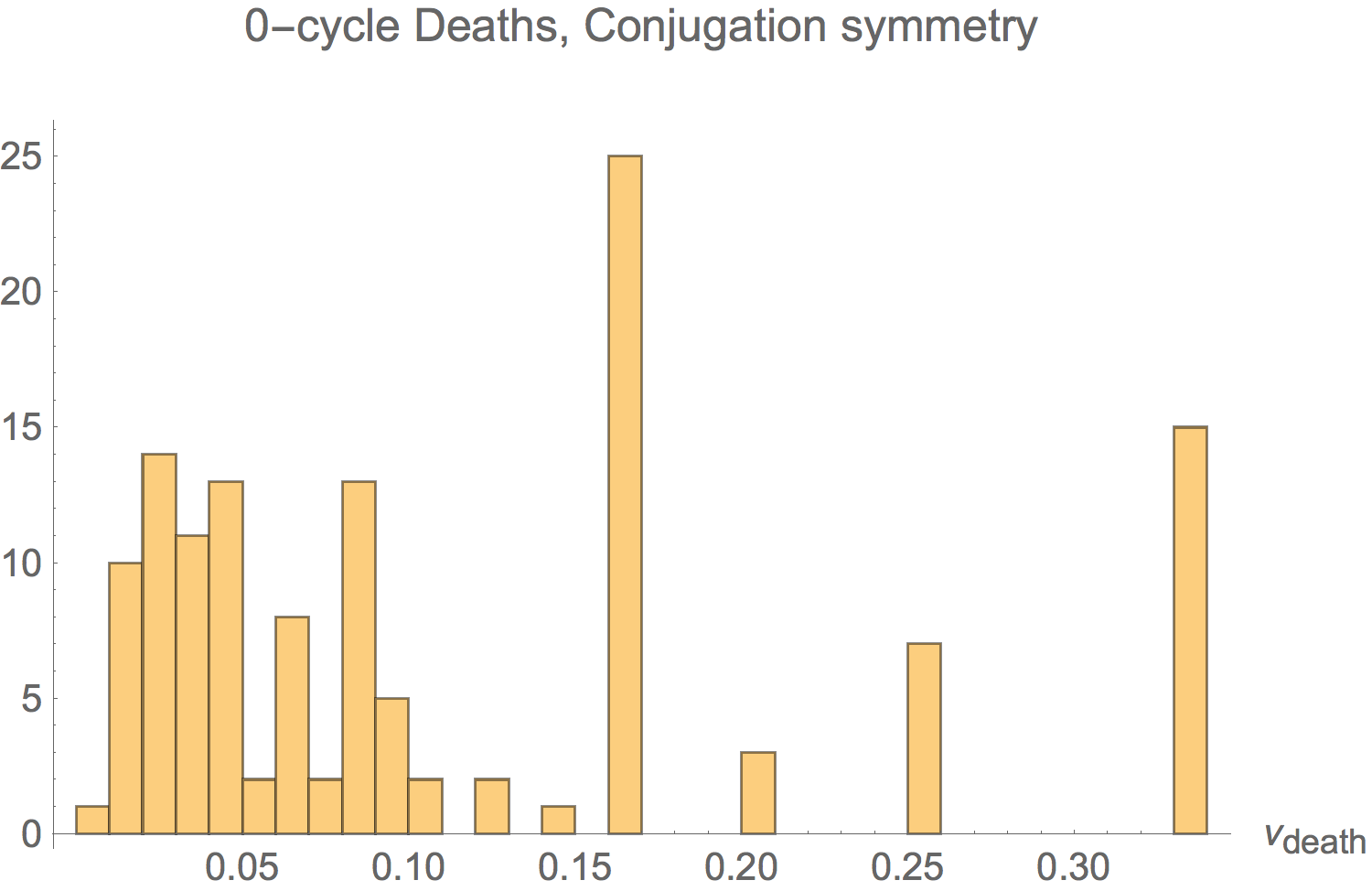}
\caption{For rigid Calabi-Yau vacua with a low energy symmetry descending from complex conjugation, the topological complexity of the distribution is reduced. As the points live on a line, there are only 0-cycles. Long-lived 0-cycles correspond to vacua at the centers of voids in the full distribution.}\label{fig:conjRigid}
\end{figure}

We can also consider discrete symmetries in this model. There is a low-energy symmetry descending from complex conjugation for vacua with imaginary axiodilaton.\footnote{There are also vacua with enhanced symmetries descending from $SL(2,\mathbb{Z})$, but as previously discussed, these only occur at isolated points in the moduli space and thus are not topologically interesting. The rigid Calabi-Yau does not have any $W=0$ vacua.} We might consider the persistent homology of the distribution of these special vacua. Restricting to the imaginary axis, we can only reduce the topological complexity of the distribution. As previously discussed, the restricted distribution cannot feature 1-cycles. All 0-cycles are born at the beginning of the filtration, so instead of a persistence diagram we show a histogram of the 0-cycle deaths in Fig. \ref{fig:conjRigid}. There are many 0-cycles that die at $\nu_{\rm death}=1$. These vacua live at large Im $\phi$, where tadpole cancellation and flux quantization dictate that the nearest neighbor should be a unit distance away. There are also a fair number of 0-cycles dying at $\nu_{\rm death}=0.5$ for similar reasons. For smaller Im $\phi$, relatively long-lived 0-cycles will correspond to vacua at the centers of voids (or rather, what would be voids in the full distribution). Thus certain aspects of the full distribution show up (albeit in the form of lower-dimensional cycles).

In this example we could have simply looked at the projection in Fig. \ref{fig:rigidAxio} and understood the presence of voids with isolated vacua in their centers. However, persistent homology also proves useful in higher-dimensional examples where we do not have a simple visualization. Moreover, we learned that persistence pairing can be used to confirm suspicions arising from correlations in persistence diagrams. We also had our first example of restricting to vacua satisfying special properties. In this case the restriction only reduced the topological complexity of the distribution.
\subsection{Calabi-Yau Hypersurface and Multiparameter Persistence}\label{sec:hyp}
In this section we consider the hypersurface defined by
\begin{align}\label{eqn:hyp}
  \sum_{i=1}^4 x_i^8+4x_0^2-8\psi x_0x_1x_2x_3x_4=0
\end{align}
in the weighted projective space ${\bf WP}^{4}_{1,1,1,1,4}$. The hypersurface has $h_{1,1}=1$ and $h_{2,1}=149$. For this hypersurface, a particular orientifold (taking $x_0\to-x_0,\psi\to-\psi$ along with worldsheet parity reversal) arises from F-theory compactified on a Calabi-Yau fourfold defined as a hypersurface in ${\bf WP}^5_{1,1,1,1,8,12}$, giving a tadpole condition $L_{\rm max}=972$ \cite{Giryavets:2003vd}. As described in \cite{Giryavets:2003vd,Giryavets:2004zr,DeWolfe:2004ns}, the hypersurface equation (\ref{eqn:hyp}) has a discrete symmetry group $\Gamma=\mathbb{Z}_8^2\times \mathbb{Z}_2$. Any deformation to the complex structure besides the $\psi$ term is charged under $\Gamma$. Thus, if only fluxes consistent with $\Gamma$ are turned on, these charged moduli can only appear at higher order in the superpotential. We can then consistently set the charged moduli to zero, solve for the periods, and compute the vevs for the axiodilaton and uncharged modulus $\psi$.

We will focus on flux vacua near the conifold point $\psi=1$. To first order in $x\equiv 1-\psi$, the F-flatness conditions give
\begin{align}\label{eqn:hyp1}
  \phi&=\frac{f_1\overline{a}_0+f_2\overline{b}_0+f_3\overline{c}_0}{h_1\overline{a}_0+h_2\overline{b}_0+h_3\overline{c}_0}+\mathcal{O}(|x|\ln|x|)\\
  \ln(x)&=-\frac{2\pi i}{d_1}\left[\frac{(f_1-\phi h_1)(a_1-\frac{\mu_1}{\mu_0}a_0)+(f_2-\phi h_2)(b_1-\frac{\mu_1}{\mu_0}b_0)}{f_2-\phi h_2}+\right.\nonumber\\
  &\left.\quad\quad\quad\quad\quad\quad \frac{(f_3-\phi h_3)(c_1-\frac{\mu_1}{\mu_0}c_0)+(f_4-\phi h_4)d_1}{f_2-\phi h_2}\right]-1\label{eqn:hyp2}
\end{align}
where constants can be found in \cite{Giryavets:2004zr}. Monte Carlo sampling in \cite{Giryavets:2004zr} explicitly confirmed the expectation from \cite{Ashok:2003gk} that vacua would cluster near the conifold point, including the scaling of density with distance from the conifold point.\footnote{One might be tempted to worry about the breakdown of our EFT here due to the masslessness of a wrapped D-brane state \cite{Strominger:1995cz}. We can consistently leave this state out of our EFT by going to larger volume, since the K\"ahler moduli are unfixed in our setup. Increasing the Calabi-Yau volume increases the wrapped D-brane's mass and lowers the masses of the moduli under consideration, which scale as the flux density. It would be interesting to see how including this state in the full moduli space modifies the clustering as calculated by the approximation of \cite{Ashok:2003gk,Denef:2004ze}.} 

We would like to study this clustering using persistent homology. In this case, since we have four real dimensions (two from $\phi$ and two from $x$), there is no simple visualization of the space of vacua, although we may plot projections onto specific planes. Although the clustering is manifest in the projection onto $x$, there could be more structure relating the clustering to $\phi$. Persistent homology in the full four-dimensional space allows us to search for higher-dimensional features in the distribution of vacua.

We also encounter a new issue. While a length-based filtration like Vietoris-Rips is well-suited for void identification, it does not perform as well in identifying clusters. Using a length-based filtration, the persistence diagram for a cluster is not very different from the persistence diagram for a uniform distribution. The cluster will give rise to many 0-cycles that die very early in the filration. This is not using persistent homology in a very clever way.

One way around this is to use a \emph{multiparameter} filtration \cite{carlsson2009theory}. In addition to the length parameter, we can assign to each point a density (e.g. the inverse distance of the $n$-th nearest neighbor). We can then take as an orthogonal filtration parameter a threshold density $\rho_{\rm th}$ and only include points with $\rho<\rho_{\rm th}$. For low $\rho_{\rm th}$, points in the cluster will not yet be included, and the distribution will feature a void, easily picked up by a length-based filtration. This density filtration is related to the sublevel filtration the present authors used to study the CMB in \cite{Cole:2017kve}. We can imagine smoothing out the point cloud to define a density function on the moduli space. We are then performing a sublevel filtration on this function. How well this represents the underlying point cloud depends on how well-sampled the space is.

While multiparameter persistence is well-defined, it lacks a simple summary statistic. Instead, one has a persistence diagram for each (well-defined) path through $(r,\rho_{\rm th})$ space. (A software implementing interactive visualization of two-dimensional persistence is \cite{2015arXiv151200180L}.) For our purposes, we will only consider a length-based filtration at two density thresholds to demonstrate the successful identification of the cluster and diagnose whether there is any more interesting structure in the four-dimensional space.

\begin{figure}
\centering
\includegraphics[width=0.5\textwidth]{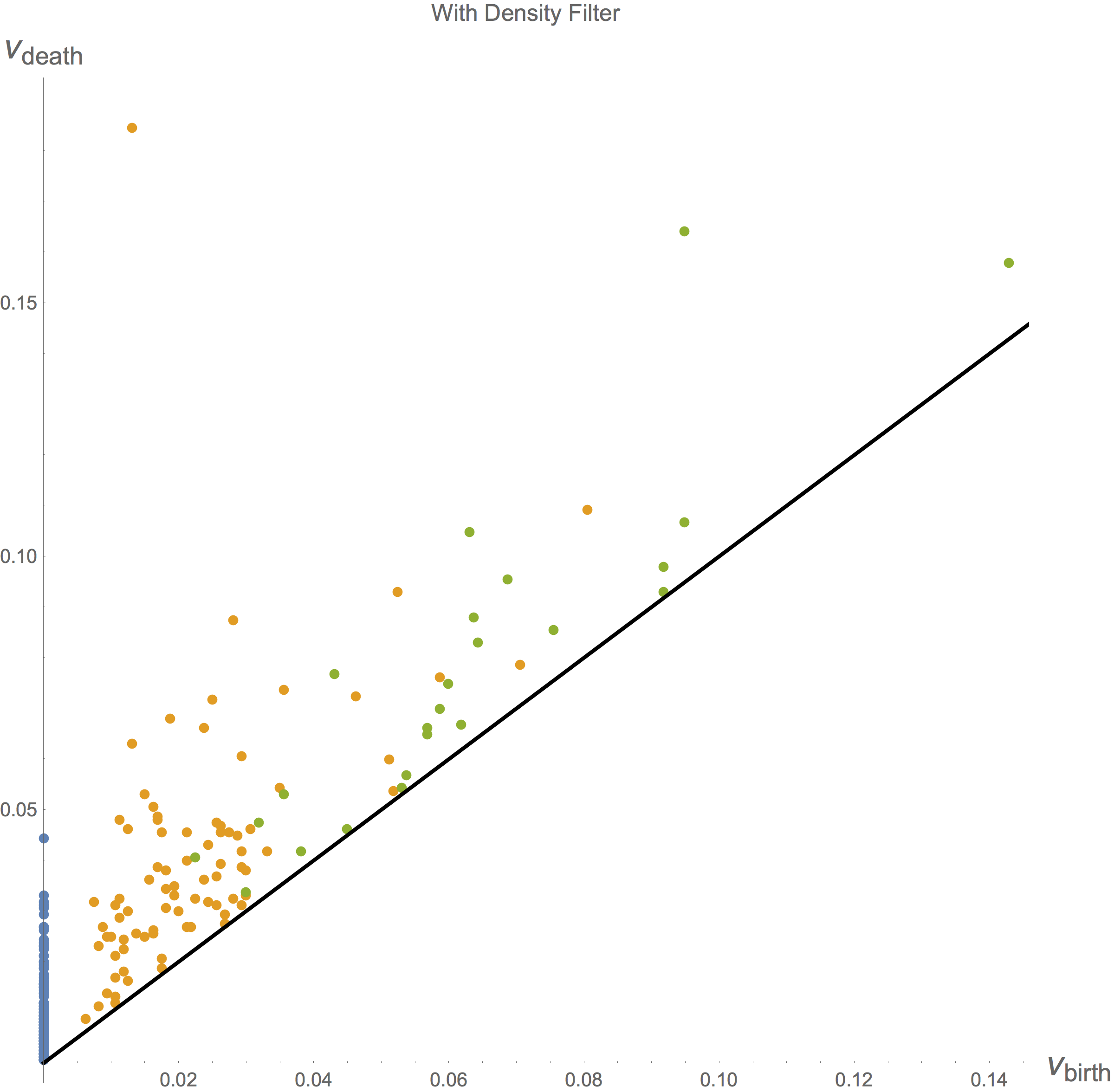}\includegraphics[width=0.5\textwidth]{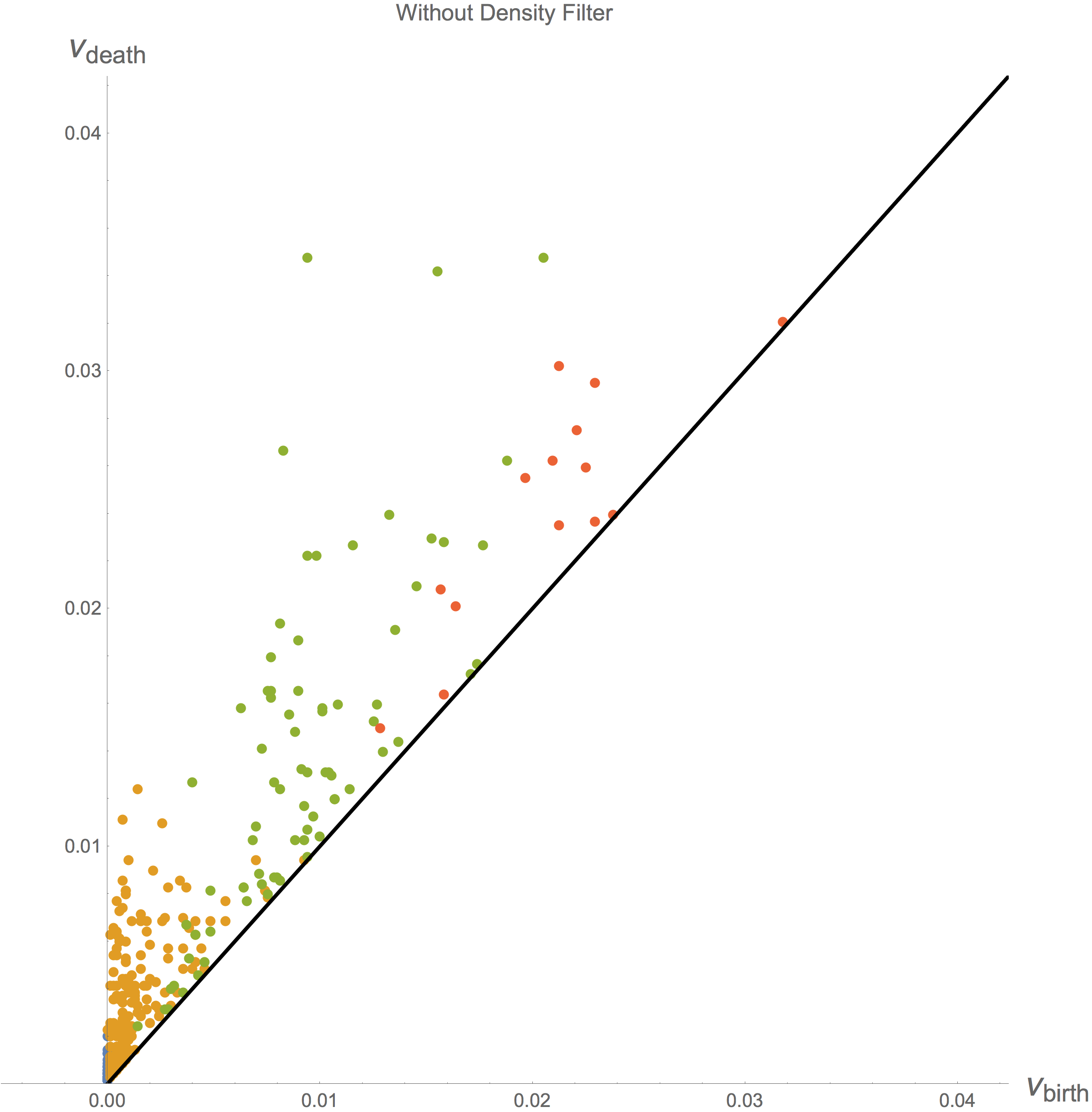}
\caption{Left: persistence diagram for lazy witness complex with 200 landmark points and density filter applied. In blue are 0-cycles, in orange 1-cycles, and in green 2-cycles. There is a long-lived 1-cycle corresponding to the excised cluster. Right: without the density filter, there is no long-lived 1-cycle. We observe a few 3-cycles (red), but, like the rest of the features, they are short-lived.}\label{fig:hypPD}
\end{figure}

We generate the point cloud by drawing random flux vectors and computing the stabilized moduli with (\ref{eqn:hyp1}),(\ref{eqn:hyp2}). We have to be careful to fix the gauge symmetry and stay within the regime of perturbative validity $|x|\ll 1$. In addition to S-duality for the axiodilaton, the complex structure modulus has a logarithmic monodromy around the conifold point. We fix the gauge by mapping $\phi$ to the $SL(2,\mathbb{Z})$ fundamental domain (also acting on the fluxes) and taking $\arg x\in[-\pi,\pi)$.

Computing the persistence diagrams for the point cloud with and without a density filter, we find a long-lived 1-cycle with the filter and no nontrivial features without the filter (Fig. \ref{fig:hypPD}). Thus we have recovered the conifold clustering using persistent homology. Moreover, we find no higher-dimensional features linking the clustering in $x$ to the distribution for $\phi$. In other words, the clustering in $x$ does not correlate with any topological structure in $\phi$. This is an aspect we could not have diagnosed with simple visualization.

\subsection{Symmetric $T^6$}
\label{sec:t6}In this section we consider toroidal compactifications where a $T^6$ can be viewed as a direct product of three two-tori with equal modular parameter $\tau$. The moduli space has two complex dimensions, so attempts at simple visualization require projecting onto arbitrary planes. Persistent homology, however, has no trouble with higher-dimensional spaces, and can be used to characterize the distribution of vacua. The symmetric $T^6$ also features vacua with vanishing tree-level superpotential. These vacua give us the opportunity to study how imposing phenomenologically desirable conditions affects the persistent homology of a distribution of vacua that cannot be directly visualized. 

We follow the conventions of \cite{DeWolfe:2004ns}. Take $x^i,y^i$ for $i=1,2,3$ to be coordinates with periodicity $x^i\equiv x^i+1,y^i\equiv y^i+1$. Then choose three holomorphic 1-forms $dz^i=dx^i+\tau^{ij}dy^j$. We take the orientation
\begin{align}
	\int dx^1\wedge dx^2\wedge dx^3\wedge dy^1\wedge dy^2\wedge dy^3=1
\end{align}
A symplectic basis for $H^3(T^6,\mathbb{Z})$ is
\begin{align}
	\alpha^0&=dx^1\wedge dx^2\wedge dx^3\nonumber\\
	\alpha_{ij}&=\frac{1}{2}\epsilon_{ilm}dx^l\wedge dx^m\wedge dy^j\nonumber\\
	\beta^{ij}&=-\frac{1}{2}\epsilon_{jlm}dy^l\wedge dy^m\wedge dx^i\nonumber\\
	\beta^0&=dy^1\wedge dy^2\wedge dy^3
\end{align}
The holomorphic 3-form is $\Omega=dz^1\wedge dz^2\wedge dz^3$. We can expand the fluxes as 
\begin{align}
	F_3&=a^0\alpha^0+a^{ij}\alpha_{ij}+b_{ij}\beta^{ij}+b_0\beta^0\nonumber\\
	H_3&=c^0\alpha^0+c^{ij}\alpha_{ij}+d_{ij}\beta^{ij}+d_0\beta^0
\end{align}
So far, we have written our parameterization for a general $T^6$. Now we specialize to the symmetric case, taking $\tau^{ij}=\tau\delta^{ij}$. In other words, we take the $T^6$ to be factorizable as three two-tori with equal modular parameter. This corresponds to turning on only a special subset of the fluxes (as in the hypersurface example):
\begin{align}
	a^{ij}=a\delta^{ij},\quad b_{ij}=b\delta_{ij},\quad c^{ij}=c\delta^{ij},\quad d_{ij}=d\delta_{ij}
\end{align}
In this case, the superpotential takes the form
\begin{align}
	W=P_1(\tau)-\phi P_2(\tau)
\end{align}
where $P_i$ are cubic polynomials in $\tau$ over the integers
\begin{align}
	P_1(\tau)&\equiv a^0\tau^3-3a\tau^2-3b\tau-b_0\\
	P_2(\tau)&\equiv c^0\tau^3-3c\tau^2-3d\tau-d_0
\end{align}
The K\"ahler potential for $\tau$ and $\phi$ is 
\begin{align}
	\mathcal{K}=-3\log(-i(\tau-\overline{\tau}))-\log(-i(\phi-\overline{\phi}))
\end{align}
and the flux-induced D3-brane charge is
\begin{align}
	N_{\rm flux}=b_0c^0-a^0d_0+3(bc-ad)
\end{align}
The F-flatness conditions can be simplified to be
\begin{align}
	P_1(\tau)-\overline{\phi}P_2(\tau)&=0\\
	P_1(\tau)-\phi P_2(\tau)&=(\tau-\overline{\tau})(P_1'(\tau)-\phi P_2'(\tau))
\end{align}
The axiodilaton can be solved for using one of these equations and plugged into the other, giving the equations for $\tau=x+iy$
\begin{align}
  q_1(x)y^2&=q_3(x)\label{eqn:t6y1}\\
  q_0(x)y^4&=q_4(x)\label{eqn:t6y2}
\end{align}
where the $q_i$ are polynomials in $x$ whose form can be found in the appendix of \cite{DeWolfe:2004ns}. When these equations are multiplied to eliminate $y$, miraculous cancellations leave one with a cubic rather than sextic equation in $x$,
\begin{align}
\alpha_3x^3+\alpha_2x^2+\alpha_1x+\alpha_0=0\label{eqn:t6x}
\end{align}
where the $\alpha_i$ are combinations of flux integers that can also be found in \cite{DeWolfe:2004ns}.
\subsubsection{Generic vacua}
\begin{figure}
\centering
\includegraphics[width=0.5\textwidth]{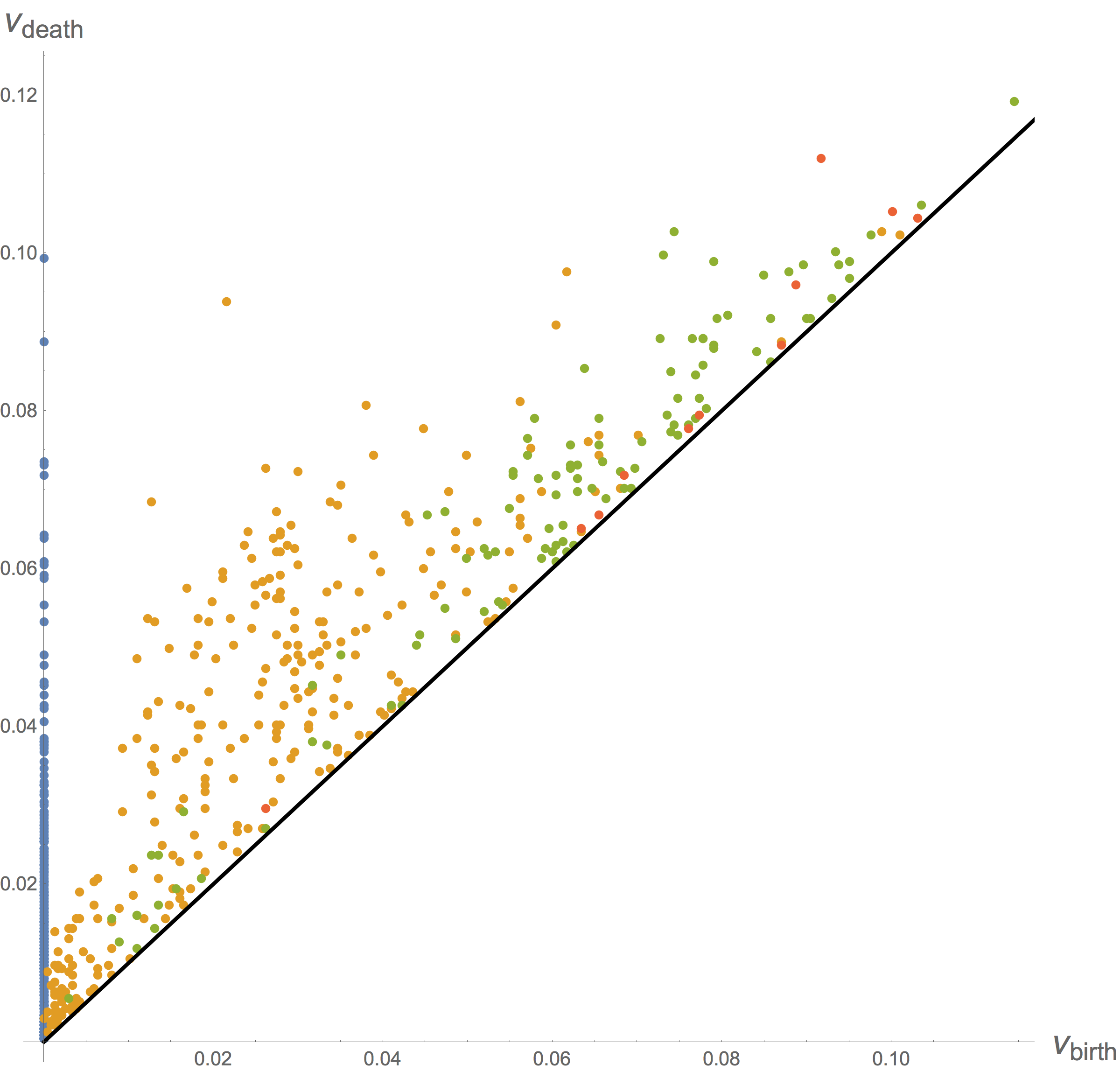}\includegraphics[width=0.5\textwidth]{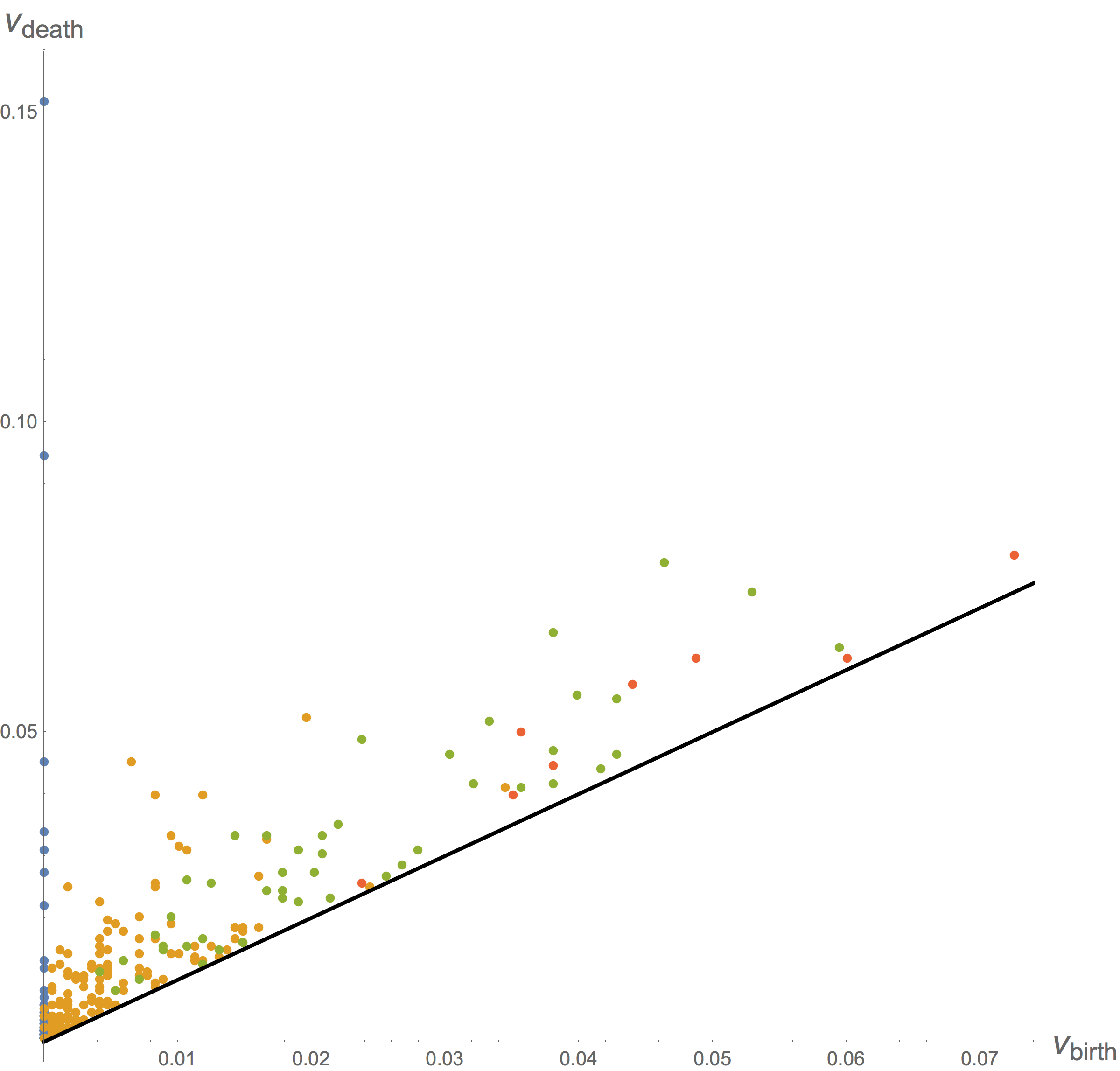}\caption{Left: persistence diagram for generic vacua and $L_{\rm max}=18$. Right: persistence diagram for generic vacua and $L_{\rm max}=54$, looking at the underdense region. We see that as $L_{\rm max}$ is increased, the higher dimension cycles become shorter-lived.}\label{fig:T6gen}
\end{figure}

\begin{figure}
\centering
\includegraphics[width=0.5\textwidth]{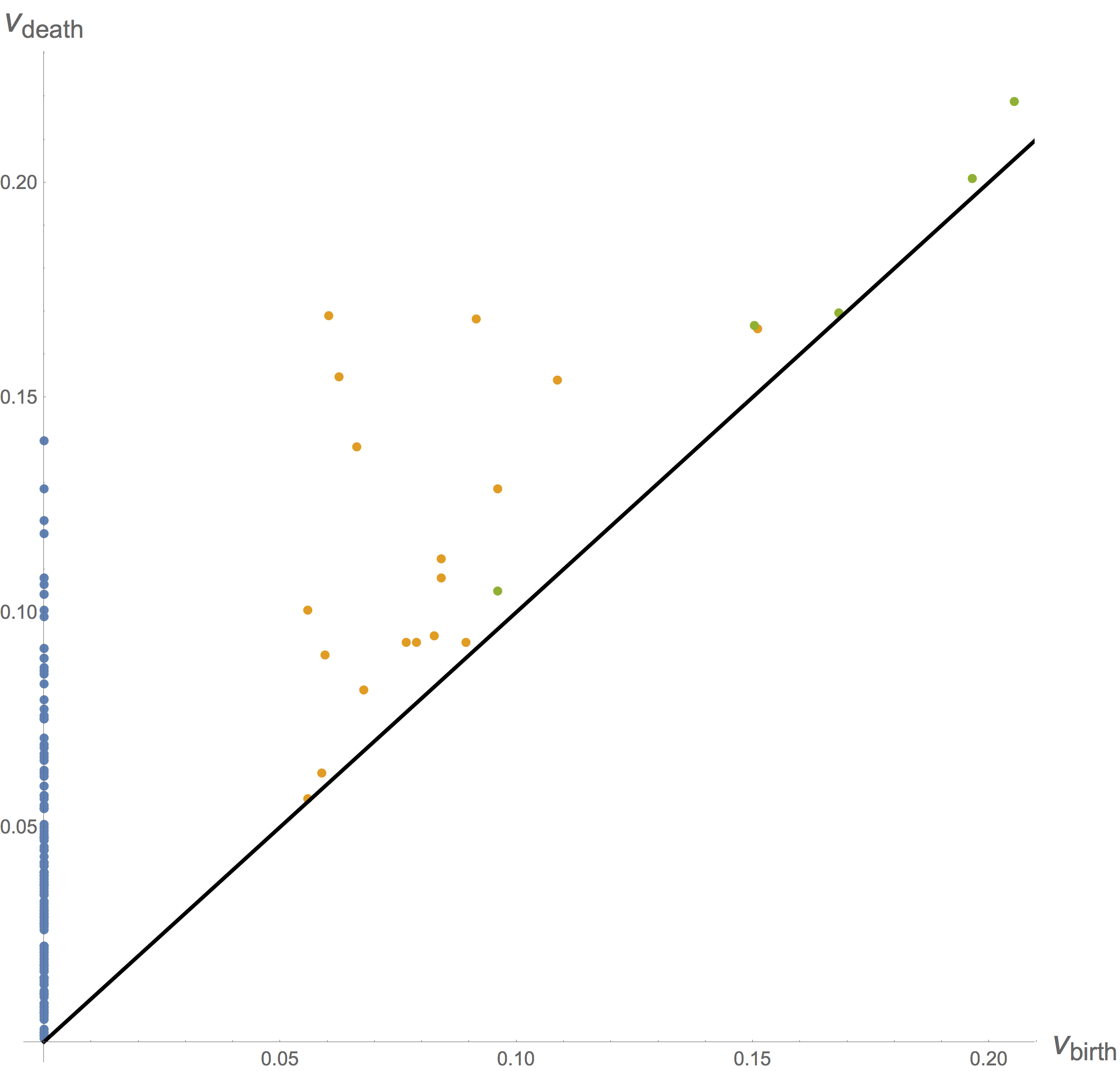}\caption{Long-lived 1-cycles in the generic distribution of $T^6$ vacua. Here we used 150 landmark points for a subregion of the generic $T^6$ distribution with $L_{\rm max}=54$.}\label{fig:t6voids}
\end{figure}

We generate random flux vectors, solve for $\tau$ and $\phi$, and map both to their fundamental domains to fix the gauge. For $L_{\rm max}\leq 18$, lattice effects suppress the number of vacua (essentially, (\ref{eqn:t6y1}) and (\ref{eqn:t6x}) are not compatible). As $L_{\rm max}$ is increased above 18, a dense region develops for small $\textrm{Im}~\tau$ with mostly trivial topology at our resolution and sampling, pushing an underdense region that is topologically similar to the $L_{\rm max}=18$ distribution to larger $\textrm{Im}~\tau$. However, it seems that this underdense region loses some topological complexity as it is pushed (Fig. \ref{fig:T6gen}). Moreover, the dense region at small $\textrm{Im}~\tau$ is not entirely trivial. Actually, voids similar to those in the rigid model develop on the imaginary axis, shrinking as $L_{\rm max}$ is increased (Fig. \ref{fig:t6voids}).

\subsubsection{$W=0$ vacua}

The symmetric $T^6$ features vacua with vanishing tree-level superpotential, $W=0$. Considering only these vacua, we find that their distribution in moduli space exhibits different topological structure than the generic vacua. In particular we find that restricting to $W=0$ vacua results in additional higher-dimensional cycles in certain regions of the distribution.

Combined with the F-flatness conditions, enforcing $W=0$ means the simultaneous vanishing of $P_1(\tau)$ and $P_2(\tau)$,
\begin{align}
	P_1(\tau)&= a^0\tau^3-3a\tau^2-3b\tau-b_0=0\\
	P_2(\tau)&= c^0\tau^3-3c\tau^2-3d\tau-d_0=0
\end{align}
For details on enumerating fluxes giving rise to $W=0$ vacua, see \cite{DeWolfe:2004ns}. The important point is that the condition $W=0$, when combined with flux quantization and tadpole cancellation, maps to different structure in the moduli space than the generic vacua exhibit. Moreover, as we will see, restricting to $W=0$ vacua, unlike the discrete symmetry restriction of Sec. \ref{sec:rigid}, induces \emph{richer} topology in the distribution. This is possible because $W=0$ is less restrictive about the stabilized vevs of the moduli.

\begin{figure}[h!]
\centering
\includegraphics[width=0.5\textwidth]{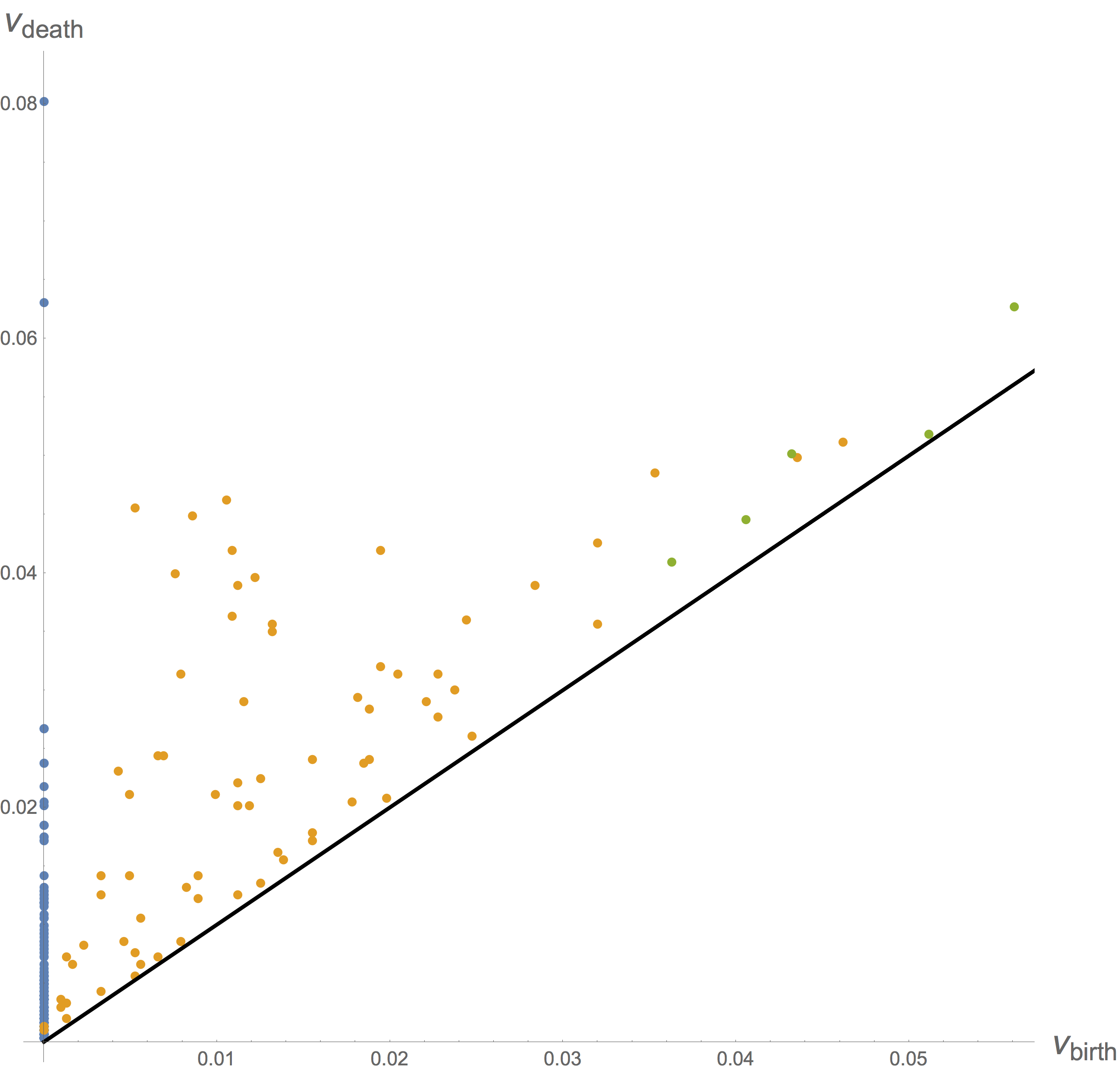}\caption{Large-scale structure of $W=0$ vacua for $L_{\rm max}=504$. No higher-dimensional cycles live past the last disconnected component of the point cloud.}\label{fig:t6wl}
\end{figure}

\begin{figure}[h!]
\centering
\includegraphics[width=0.5\textwidth]{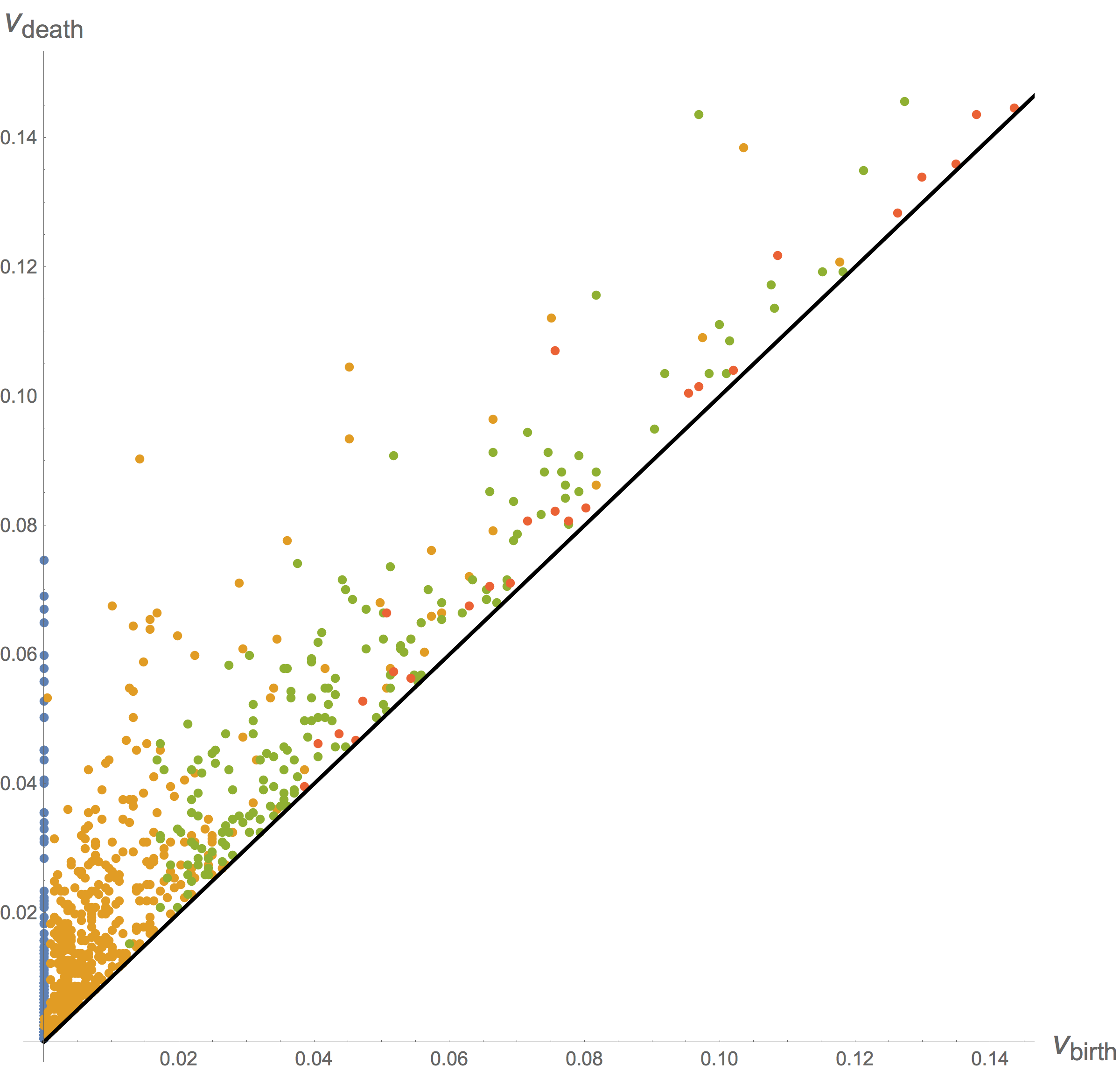}\caption{Small-scale structure of $W=0$ vacua. Long-lived higher-dimensional cycles are present in the distribution well after there is only one connected component left. Restricting to $W=0$ vacua increases the topological complexity of the distribution.}\label{fig:t6ws}
\end{figure}

Given $L_{\rm max}$, we generate all $W=0$ vacua with $N_{\rm flux}\leq L_{\rm max}$. We again fix gauge by mapping to the fundamental domain. We observe a large-scale structure that is insensitive to $L_{\rm max}$ (Fig. \ref{fig:t6wl}). Moreover, for sufficient $L_{\rm max}$, there are enough $W=0$ vacua to form a complex topology in moduli space (Fig. \ref{fig:t6ws}). The multiscale topology of such vacua contains higher-dimensional cycles that are long-lived. In this case, we see that restricting to $W=0$ vacua gives rise to a more complex topology. Not only is the topology more complex than the dense subregion of the generic distribution in which we are making our cut, but it is also more complex than the underdense subregions of the distribution (cf. Fig. \ref{fig:T6gen}). This is perhaps to be expected, as the extra requirement $W=0$ imposes richer number-theoretic structure on the flux space.

\section{Conclusion}\label{sec:conc}
Persistent homology can be used to study the shape of a data set. For string theory, we are interested in understanding the structure of the landscape. We showed that persistent homology can be used to effectively characterize generic and special vacua in toy models. Despite the simplicity of our toy examples, we learned a few things that should prove useful in a scaled-up program. For one, we learned that persistence pairing can be used to recover more refined information than is expressed in a persistence diagram. In particular, we were able to reconstruct the presence of isolated vacua inside voids in the rigid Calabi-Yau construction. We also learned that to robustly characterize not only underdense regions like voids but also overdense regions like clusters, a notion of multiparameter persistence is useful. In addition to the usual length-based filtration parameter, one can consider an orthogonal density threshold parameter. As the density threshold is raised, we include points in dense regions, allowing persistent homology to recognize the presence of clusters by their excision. This notion could have interesting cosmological applications, e.g., in the context of  \cite{2011MNRAS.414..350S,2011MNRAS.414..384S,Pranav:2016gwr,Cole:2017kve,Xu:2018xnz,Elbers:2018fus}. We also studied the persistent homology of restricted distributions of special vacua (like those with discrete low-energy symmetries or vanishing tree-level superpotential). Understanding the topology of special vacua could have interesting consequences in more realistic models if we want to ask about the simultaneous satisfiability of multiple low-energy criteria and the distribution of those \emph{very} special vacua.

There are plenty of future directions to consider. Obviously, we would like to study distributions arising from more realistic constructions. In this work we only considered the distribution of complex structure moduli and axiodilaton vevs stabilized by fluxes is type IIB string theory. In principle we also know how to stabilize the K\"ahler moduli and open string moduli. We also largely ignored the fluxes. Trying to combine the fluxes with the moduli vevs seemingly presents a problem, since the fluxes are discrete while the moduli vevs in our cases were rational or irrational. However, the fluxes do show up in the low-energy theory as coupling constants, and we could include them in a unified analysis by considering e.g. the masses of stabilized moduli.

Of course, as we move on to more realistic data sets, the constructions necessarily become more complicated. While we made use of persistent homology's ability to compute in high dimensions, we did not consider models with hundreds of moduli. Understanding how to adapt our techniques to such situations might require advances on the algorithmic/software side (see \cite{2015arXiv150608903O} for a comparison of different software packages). Another difficulty is in choosing parameters (number of witness points, subregion of moduli space, maximum filtration parameter) for the persistent homology computation. For this paper, we largely chose parameters via a ``guess and check'' method, as certain parameter values would e.g.\ freeze the program or not find what we were looking for. Ultimately we would like to perform a systematic scan over a large database of string models. Automating the choice of parameters then presents another difficulty. With such an automated scan, we would also have the problem of too much output. Systematically processing persistence diagrams provides another challenge. It could be here that persistent homology and machine learning techniques might be usefully coupled.

From a more physical perspective, it would interesting to ask what structure in the landscape means for vacuum selection and tunneling between vacua. This is where related ideas in the study of energy landscapes may become useful. Morse theory relates the topology of sublevel sets with critical points of a potential. Therefore, TDA can be used to sample the topology of the string landscape to effectively find vacua and study their transitions. We plan to return to this interesting idea in a future work. An alternative approach to these questions from the perspective of network science was advocated in \cite{Carifio:2017nyb}. However, much of the interesting dynamics is suppressed in treating string vacua as nodes of a network. More understanding is also needed on the physics side here to study the transitions. For example, tunneling in the presence of dynamics remains poorly understood \cite{Brown:2017wpl}.

In this paper we were concerned with methods for mapping out the structure of the landscape itself. A complementary approach is to study the even vaster \emph{swampland} of effective field theories with gravity that do not admit UV completions \cite{Vafa:2005ui}. 
Recent conjectures put some interesting constraints on the shape of the energy landscape of string theory \cite{Obied:2018sgi,Ooguri:2018wrx}.
It would also be interesting to apply TDA or other data science techniques to understand the shape of the boundary between the landscape and the swampland, or to test various conjectures about the topology of moduli space \cite{Ooguri:2006in}.
 
\subsection*{Acknowledgments}
We used the publicly available package Javaplex \cite{Javaplex}, modified to extract persistence pairs, for our persistent homology calculations. We would like to thank Jon Brown and Aitor Landete for helpful discussions.
This work is supported in part by the DOE grant DE-SC0017647 and the Kellett Award of the University of Wisconsin.

\bibliographystyle{utphys}\bibliography{TDA_Vacua.bib}

\end{document}